\documentclass[twocolumn,prl]{revtex4}

\usepackage{graphicx}
\usepackage{dcolumn}
\usepackage{bm}
\usepackage{multirow}
\usepackage{color}
\usepackage[normalem]{ulem}
\usepackage{amsmath}
\usepackage{gensymb}
\usepackage[colorlinks=true]{hyperref}

\definecolor{mblue}{rgb}{0,0.35,0.75}
\definecolor{mgreen}{rgb}{0,0.5,0.5}

\begin{document}
\title{Control of the Exciton Radiative Lifetime in van der Waals Heterostructures}

\author{H.H. Fang$^{1*}$}
\author{B. Han$^{1*}$}
\author{C. Robert$^1$}
\author{M.A. Semina$^2$}
\author{D. Lagarde$^1$}
\author{E. Courtade$^1$}
\author{T.~Taniguchi$^3$}
\author{K. Watanabe$^3$}
\author{T. Amand$^1$}
\author{B. Urbaszek$^1$}
\author{M.M. Glazov$^2$}
\author{X. Marie$^1$}

\affiliation{%
$^1$Universit\'e de Toulouse, INSA-CNRS-UPS, LPCNO, 135 Av. Rangueil, 31077 Toulouse, France}
\affiliation{$^2$Ioffe Institute, 194021 St. Petersburg, Russia}
\affiliation{$^3$National Institute for Materials Science, Tsukuba, Ibaraki 305-0044, Japan}

\begin{abstract}
Optical properties of atomically thin transition metal dichalcogenides are controlled by robust excitons characterized by a very large oscillator strength.  Encapsulation of monolayers such as MoSe$_2$ in hexagonal boron nitride (hBN) yields narrow optical transitions  approaching the homogenous exciton linewidth. We demonstrate that the exciton radiative rate in these van der Waals heterostructures can be tailored by a simple change of the hBN encapsulation layer thickness as a consequence of the Purcell effect. The time-resolved photoluminescence measurements show that the neutral exciton spontaneous emission time can be tuned by one order of magnitude depending on the thickness of the surrounding hBN layers. The inhibition of the radiative recombination can yield spontaneous emission time up to $10$~ps. These results are in very good agreement with the calculated recombination rate in the weak exciton-photon coupling regime. The analysis shows that we are also able to observe a sizeable enhancement of the exciton radiative decay rate. Understanding the role of these electrodynamical effects allow us to elucidate the complex dynamics of relaxation and recombination for both neutral and charged excitons.
\end{abstract}

\maketitle

The control of the spontaneous emission using a cavity to tune the number of electromagnetic modes coupled to the emitter has been demonstrated in various atomic and solid-state systems, following the pioneering work of Purcell \cite{purcell1946purcell,kleppner1981inhibited,jhe1987suppression,benisty1998impact,gerard1998enhanced,bayer2001inhibition}. Remarkably, it was shown recently that ultra-thin semiconductors such as transition metal dichalcogenide (TMD) monolayers encapsulated in hexagonal boron nitride (hBN) exhibit spontaneous emission-dominated optical transition linewidths~\cite{Cadiz:2017a,scuri2018large,back2018realization}. A very strong light matter interaction in these 2D materials has triggered a great interest both from a fundamental point of view and for possible optoelectronic applications 
\cite{Mak:2013a,park2018radiative,wang2017plane,selig2016excitonic,smolenski2016tuning,dery2016theory,dey2017gate,hong2014ultrafast,koppens2014photodetectors}. In order to enhance the optical emission, the TMD monolayers have been integrated with various photonic crystal structures \cite{galfsky2016broadband,lee2016single,chang2018crossover}.  The optical properties are governed here by very robust excitons with binding energies of a few hundreds of meV and very large oscillator strength  \cite{Wang:2018a}. Owing to hBN induced surface protection and substrate flatness which reduce the inhomogeneous broadening \cite{Cadiz:2017a}, the exciton lines in encapsulated TMD monolayers (ML) are mainly dominated by homogeneous broadening which allow for instance the realisation of very efficient atomically thin mirrors \cite{scuri2018large,back2018realization}. In these van der Waals heterostructures, the surrounding hBN layers change the dielectric environment for the excitons in the TMD monolayer, resulting in different binding energies and oscillator strengths \cite{stier2018magnetooptics,robert2018optical}. However its impact on the exciton radiative recombination dynamics due to modification of photon modes in these atomically flat layers has not been evidenced so far.

In this Letter we demonstrate that the top and bottom hBN encapsulation layers form a microcavity-like structure which controls the exciton radiative lifetime in the MoSe$_2$ monolayer through the Purcell effect. In this weak coupling regime, the escape time of spontaneous photons out of our open cavity-like structure is much shorter than the radiative lifetime and reabsorption is negligible. This is in contrast with the strong coupling regime obtained with much more reflective mirrors resulting in microcavity polaritons \cite{schneider2018two}. 
As the spontaneous emission probability is proportional to the amplitude of the electromagnetic field mode, the variation of the local density of optical modes within the cavity is at the origin of the variation of the radiative recombination rate. In time-resolved photoluminescence (PL) measurements we demonstrate that the exciton radiative lifetime in MoSe$_2$ monolayer can be tuned by about one order of magnitude as a function of the hBN thickness, in very good agreement with the calculated dependence using transfer matrix techniques  \cite{robert2018optical}. Remarkably the measured variations of the radiative lifetime measured here (typically from 1 to 10 ps) are much larger than the ones reported previously in open semiconductor cavities based on dielectric mirrors \cite{tanaka1995cavity,abram1998spontaneous}. 
The tuning of the radiative lifetime demonstrated here for encapsulated MoSe$_2$ monolayers should also apply to other semiconductor 2D materials and associated heterostructures.

\begin{figure}[t]
\includegraphics[width=0.49\textwidth,keepaspectratio=true]{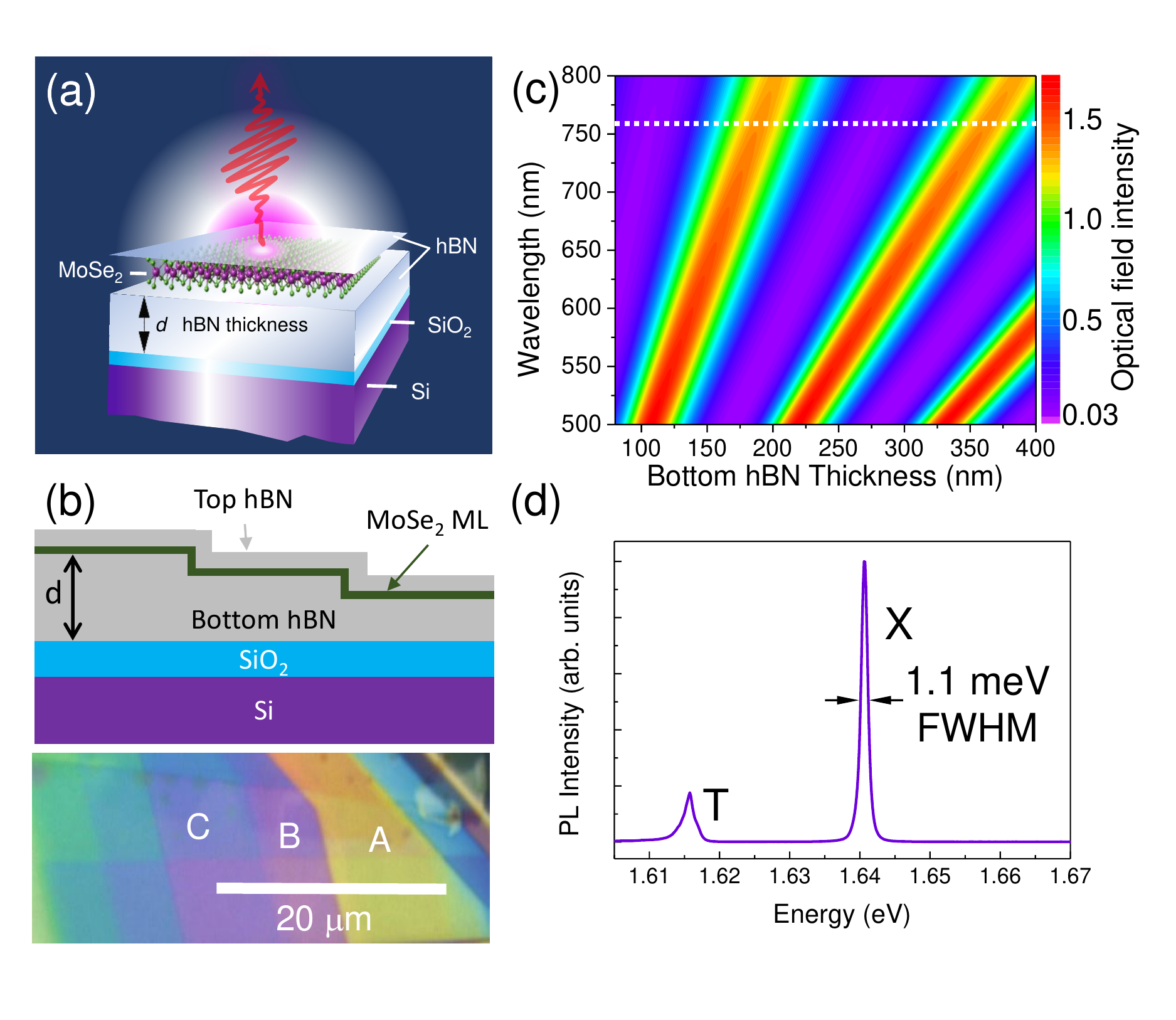}
\caption{(a) Schematics of the investigated MoSe$_2$ monolayer embedded in hexagonal Boron Nitride. (b) Schematics of the cross section and optical microscope image of the van der Waals heterostructure hBN/ML MoS$_2$/hBN (Sample III) where the same monolayer is embedded in a cavity-like structure characterized by different bottom hBN layer thickness $d$.
(c) Optical intensity map calculated at the MoSe$_2$ monolayer location as a function of both the emission wavelength and the bottom hBN layer thickness $d$. The horizontal white dotted line corresponds to the neutral exciton emission wavelength ($\sim756$  nm).
(d) \emph{cw} photoluminescence spectrum of sample II ($d=273$~nm) showing the emission of both the neutral (X) and charged (T) exciton, T=7 K.
 }
\label{fig:fig1} 
\end{figure}

\indent 
\textbf{Samples and setup.} We have investigated MoSe$_2$ MLs encapsulated in hBN deposited onto a 80 nm  SiO$_2$/Si substrate using a dry-stamping technique~\cite{castellanos2014deterministic}, see  Fig.~\ref{fig:fig1}(a) and Supplementary Information (SI~\cite{suppl1}) for the details on the fabrication technique. This easy and versatile technique allows us to fabricate various van der Waals heterostructures where the density of optical modes at the location of the TMD monolayer is tuned. During the fabrication process the thickness for each hBN layer was accurately measured by Atomic Force Microscopy (AFM) with a typical resolution of  $\pm 3$ nm for the top hBN and  $\pm 5$ nm for the bottom hBN layer. We present the results on four samples  with different bottom hBN thicknesses: In samples I and II, the bottom hBN thickness is $d$=180 and 273 nm respectively, corresponding to the MoSe$_2$ ML located, respectively, at the anti-node and the node of the  standing wave according to the calculation of the electric field distribution, Fig.~\ref{fig:fig1}(c). For the sample III, the same MoSe$_2$ ML is deposited on a hBN flake exhibiting different terraces and steps with hBN thicknesses $d=206$, 237, 247 and 358 nm for zone A, B, C and D respectively, Fig.~\ref{fig:fig1}(b) (the terrace D is outside the optical microscope image). Sample IV is similar to sample III with two terraces $d=125$ and $d=149$ nm. This allows us to investigate the exciton dynamics of the same MoSe$_2$ ML and different bottom hBN layer thicknesses. The top hBN thickness does not play a key role here considering its small value of 9, 7, 8 and 8.5 nm in sample I, II, III and IV respectively.

Figure~\ref{fig:fig1}(b) shows an optical microscope image of the Sample III illuminated with white light from a halogen lamp. For each hBN thickness, the observed color in each zone on the sample agrees very well with the one obtained by calculating the reflectivity spectra using a transfer matrix method   \cite{robert2018optical} with no adjustable parameters, using the hBN thicknesses measured by AFM and the measured hBN refractive index from Ref.~\cite{lee2018refractive}, (see SI~\cite{suppl1}).  Figure~\ref{fig:fig1}(c) presents the light intensity map calculated at the ML location as a function of both the emission wavelength and the bottom hBN thickness. The Fabry-Perot interference effects and its dependence on the bottom hBN thickness are clearly seen. 
Continuous wave (\emph{cw}) and time-resolved PL experiments are performed at $T=7$ K using a He-Ne laser (633 nm) and a Ti:Sa mode-locked laser ($\sim1.5$ ps pulse width, 80 MHz repetition rate) respectively, see the experimental details in SI~\cite{suppl1,Lagarde:2014a,Robert:2016a}. The typical excitation power is 5 $\mu$W and the spot diameter about 1 $\mu$m, i.e., in the linear regime of excitation which allows discarding any Auger type or stimulated emission processes~\cite{Chernikov:2015z}.

\indent \textbf{Results and discussion.} The encapsulation of TMD monolayers with hBN results in high optical quality samples with well-defined optical transitions exhibiting  linewidth in the 1 \ldots 4  meV range at low temperature \cite{Cadiz:2017a,wang2017probing,jin2017interlayer}. Figure~\ref{fig:fig1}(d) displays the \emph{cw} PL spectrum for sample II. In agreement with previous studies, both neutral exciton (X) and trion\, i.e., charged exciton (T) are clearly observed, with a PL linewidth of X as small as 1.1 meV (Full Width at Half Maximum, FWHM).

\begin{figure}[h]
\includegraphics[width=0.49\textwidth,keepaspectratio=true]{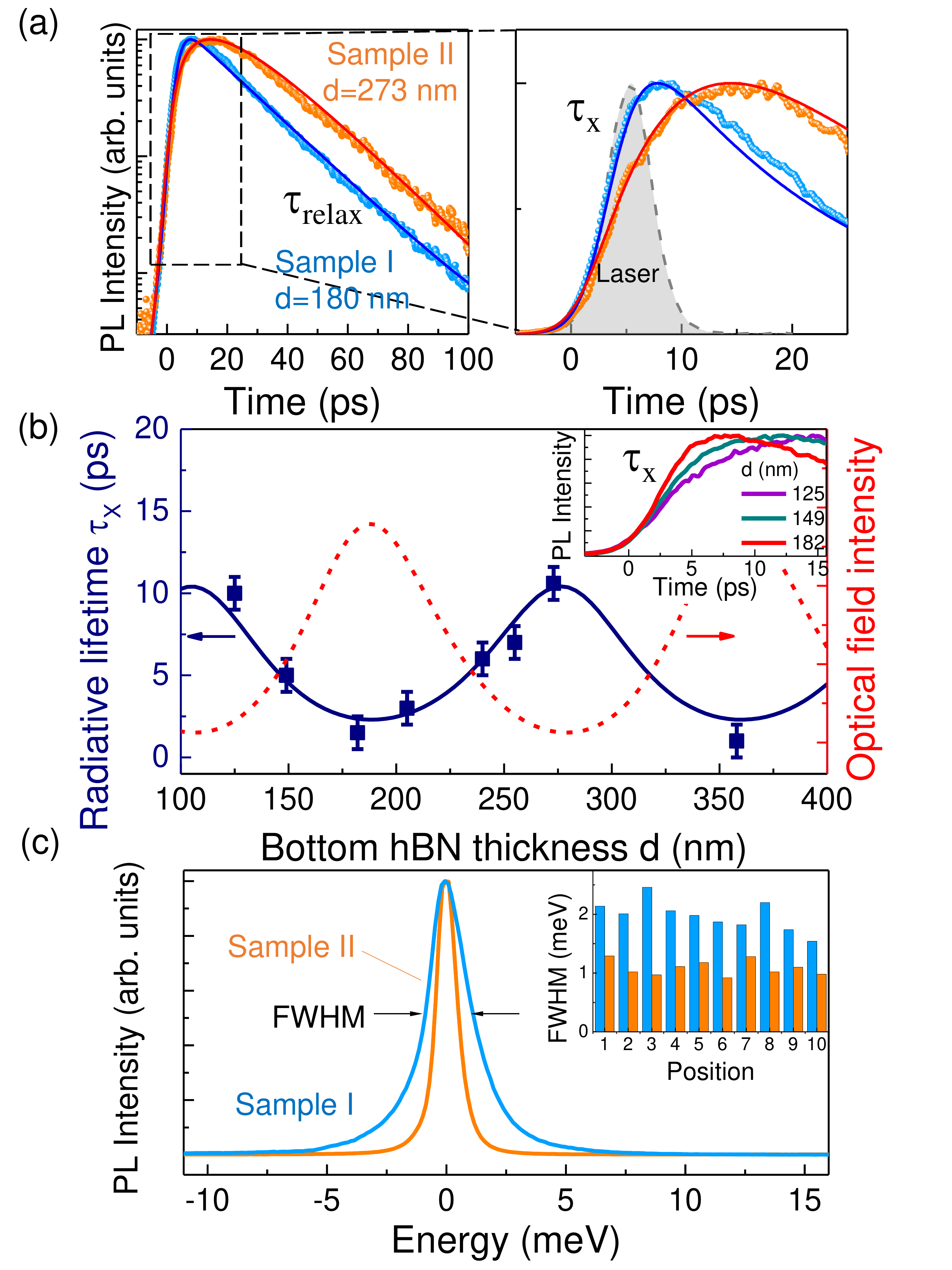}
\caption{(a) Left: normalized photoluminescence intensity (log scale) of the neutral exciton X as a function of time for sample I ($d=180$~nm) and sample II ($d= 273$~nm); the full lines correspond to the bi-exponential fits (see text). The instrument response is obtained by detecting the backscattered laser pulse (wavelength 712 nm) on the sample surface, see the dotted line labeled 'Laser'; Right: zoom of the rise-time (linear scale).
(b) Calculated (full line) and measured (symbols) neutral exciton radiative lifetime as a function of the hBN bottom layer thickness $d$. The red dashed curve is the calculated intensity of electromagnetic field in our structure (same calculation as in Fig.~\ref{fig:fig1}(c)) . Inset: normalized time-resolved photoluminescence intensity in sample III for three different hBN bottom layer thicknesses.
(c) Normalized \emph{cw} PL intensity of the neutral exciton in sample I and sample II clearly showing different linewidths. Because the energy of the PL peak slightly depends on the sample and sample' position by a few meV, the origin of the energy axis is taken at the PL peak. Inset: PL linewidth (FWHM) for 10 different positions in sample I and II.}
\label{fig:fig2} 
\end{figure}

\begin{figure}[t]
\includegraphics[width=0.49\textwidth,keepaspectratio=true]{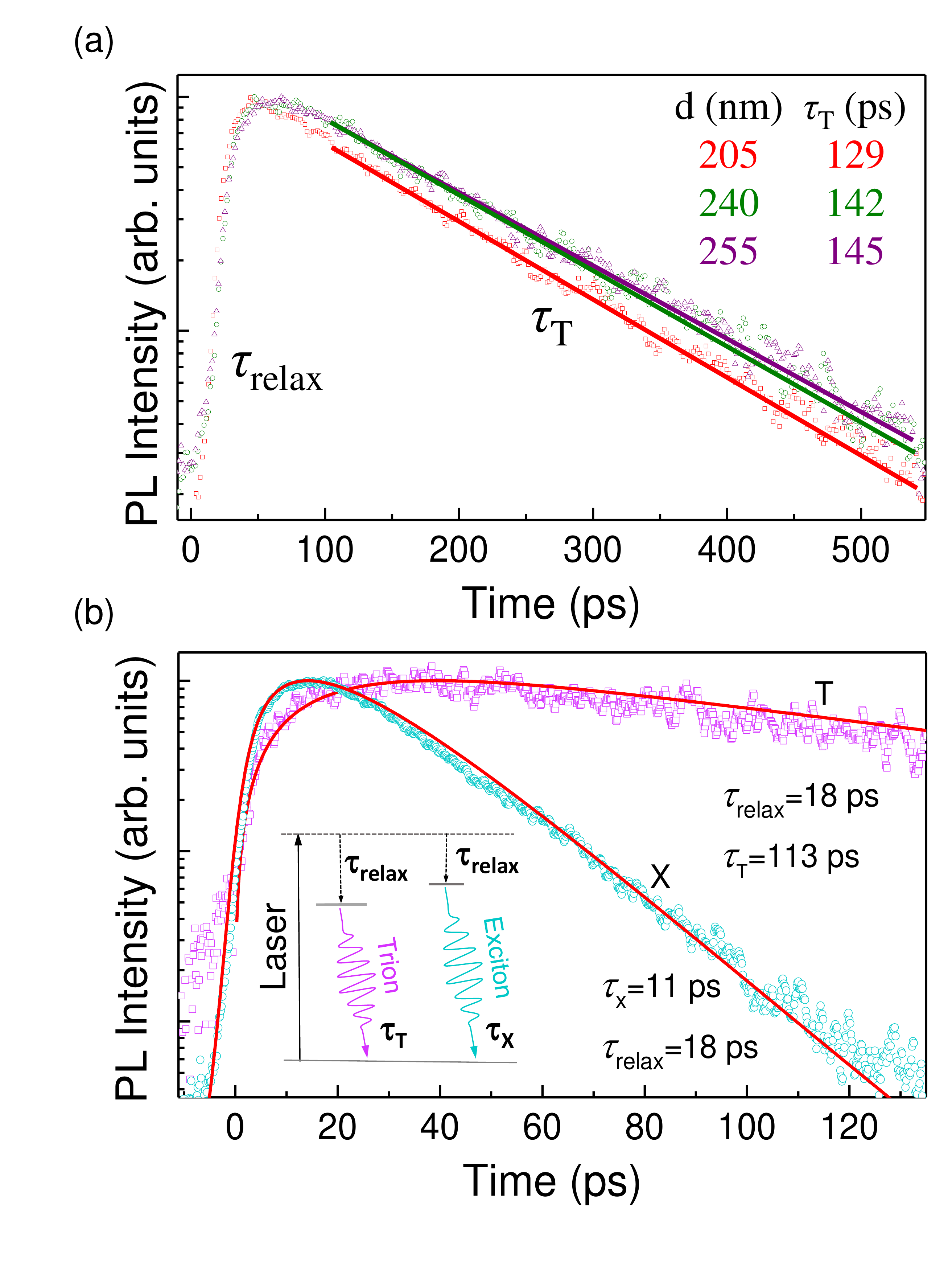}
\caption{(a) Normalized photoluminescence intensity of the charged exciton (T) as a function of time for different bottom hBN layer thicknesses $d$. The full lines correspond to mono-exponential fits of the decay time $\tau_T$.
(b) Measured (symbols) and fitted (full line) of neutral (X) and charged (T) exciton dynamics for encapsulated MoSe$_2$ monolayer with a bottom hBN layer thickness $d=273$~nm (Sample II). Inset: schematics of the two-levels model used to describe both neutral (X) and charged (T) exciton dynamics (see text).
}
\label{fig:fig3} 
\end{figure}

Figure~\ref{fig:fig2} presents the key results of this investigation. In Fig.~\ref{fig:fig2}(a), the normalized luminescence intensity dynamics of the neutral exciton X is plotted for samples I and II (differing only by the bottom hBN thickness of 180 and 273 nm respectively). While the decay time is similar in both samples with a typical value of $\sim18$ ps, the PL rise time is clearly different: it is much shorter in sample I (limited by the time-resolution of the set-up), compared to a value of $\sim 10$~ps in sample II.  In general, the rise and decay rates of PL signal are determined by the interplay between the feeding rate of the radiative state and the recombination rate. In our case, the rise time of luminescence corresponds to the exciton radiative recombination time whereas the PL decay reflects the relaxation time of photogenerated excitons at higher energies towards the radiative states ($K \approx0$). This counter-intuitive result is in part because the relaxation time, $\tau_{relax}$, is longer than recombination time, $\tau_X$, and can be easily modeled with a basic two-level model as shown in the inset of Fig.~\ref{fig:fig3}(b). The experimental results in Fig.~\ref{fig:fig2}(a) can be perfectly fitted by the resulting bi-exponential dynamics (see SI~\cite{suppl1} for details): The PL decay time is not controlled by the radiative recombination time but it corresponds to the feeding time of the radiative states, see Fig.~\ref{fig:fig3}(b) for the fit on sample II. Taking into account the instrument response time, we find $\tau_{relax}$ =18 ps in both samples whereas $\tau_X =11\pm1$ ps is typically 10 times larger in sample II compared to sample I with $\tau_X < 1.5$~ps. This is exactly the expected behaviour due to the inhibition of the spontaneous lifetime in sample II as the ML is located at the node of the electric field in the cavity-like structure (see Fig.~\ref{fig:fig2}(b)). Changing the excitation laser wavelength over the range 710-753 nm produces non measurable variations of the exciton dynamics (see SI~\cite{suppl1}). Note that in previous measurements of the exciton dynamics in bare TMDC monolayers the radiative recombination time was assigned to the decay of the emission signal \cite{Korn:2011a,Lagarde:2014a,Robert:2016a}. This control of the radiative lifetime by the cavity effect is confirmed by the measurement of the excitonic dynamics in samples III and IV where the same MoSe$_2$ monolayer is encapsulated by hBN of different thickness. Figure~\ref{fig:fig2}(b) displays the exciton radiative lifetime as a function of the hBN thickness (obtained with the same fitting procedure as above). The inset of Fig.~\ref{fig:fig2}(b) shows the measured PL rise times in sample III for different thicknesses. We have compared the measured variation with the calculated one using the transfer matrix method (see Ref.~\cite{robert2018optical} and SI~\cite{suppl1}), extracting the exciton radiative decay rate $\Gamma_0^{\rm eff}$ from the pole of numerically calculated absorbance and using the relation~\cite{Ivchenko:2005a} 
\begin{equation}
\label{radiative}
\tau_X = \hbar/(2\Gamma_0^{\rm eff}).
\end{equation}
 Assuming a free space radiative lifetime of MoSe$_2$ ML  of $2.7$~ps which is the single free parameter, we find in Fig.~\ref{fig:fig2}(b) that  the measured radiative lifetime is in very good agreement with  the calculated one. Fig.~\ref{fig:fig2}(b) demonstrates that the exciton spontaneous lifetime can be tuned by  more than one order of magnitude. This is much larger than the small variations (10-30 \% typically) reported previously with Bragg reflector microcavities using III-V semiconductor quantum wells as emitters \cite{tanaka1995cavity,abram1998spontaneous}. Significant modulations of the radiative lifetimes due to Purcell effect were evidenced in open cavities using metallic mirrors \cite{bourdon2000room} or with 3D cavity with additional lateral mode confinement: a typical factor $10$ was for instance reported for quantum dots embedded in micro-pillars \cite{gerard1998enhanced,bayer2001inhibition}. We emphasize that the radiative lifetimes in the picosecond range evidenced in Fig.~\ref{fig:fig2}  are fully consistent with the recent measurements by Four-Wave Mixing (FWM) experiments of the radiative broadening in a MoSe$_2$ monolayer encapsulated in hBN~\cite{martin2018encapsulation}.

A striking feature is that the cavity effect related to the hBN encapsulation has also a strong influence on the excitonic linewidth measured in \emph{cw} PL spectroscopy. As shown in Fig.~\ref{fig:fig2}(c), the \emph{cw}  PL linewidth is about twice smaller in sample II ($\sim1.1$ meV FWHM) compared to the one in sample I ($\sim2.2$ meV), a trend fully consistent with the expected variation of the radiative linewidth, Eq.~\eqref{radiative}, due to the cavity effect. The linewidth usually includes both a homogeneous and inhomogeneous contribution and the latter can fluctuate in different points of a given monolayer as a result of the local dielectric disorder. Nevertheless, the average of the measurements recorded for different points on the sample II (with longer $\tau_X$) is significantly lower than that on sample I. From the measurements on 10 different points on each sample, inset of Fig.~\ref{fig:fig2}(c), we find a  linewidth (FWHM) of  1.1 $\pm 0.13$ meV and 2.0$\pm 0.25$ meV on sample II and I respectively. As expected a larger linewidth is measured in sample I characterized by a much shorter radiative lifetime, see Fig.~\ref{fig:fig2}(a). This result is confirmed for sample IV for different cavity lengths (see SI).
As the exciton linewidth in TMDC monolayers is mainly dominated by radiative broadening \cite{Moody:2015a,Cadiz:2017a,jakubczyk2016radiatively,scuri2018large,back2018realization}, the control of the exciton spontaneous lifetime due to the cavity effect evidenced in Fig.~\ref{fig:fig2}(a) and (b) also yields a tuning of the exciton linewidth \cite{ zhou2019controlling,rogers2019coherent}. However, linear techniques such as photoluminescence or reflectivity spectroscopy used here cannot disentangle the linewidth contributions from inhomogeneous broadening, non-radiative processes, light scattering and radiative decay. This would require the use of non-linear techniques such as FWM experiments  \cite{Moody:2015a,jakubczyk2016radiatively,martin2018encapsulation}. Nevertheless, the exciton linewidth measured in sample I allows us to estimate the radiative lifetime in this sample (the time-resolved PL measurements demonstrate that it is shorter than $\sim 1.5 $ ps); from the analysis presented in the SI~\cite{suppl1} we can infer $\tau_X \sim740$ fs. This value is close to previous estimations where the cavity effect was not considered \cite{jakubczyk2016radiatively,martin2018encapsulation,scuri2018large,back2018realization,palummo2015exciton}. By comparing the measured radiative lifetime and the measured linewidth in \emph{cw} PL, we find that the latter is not fully controlled by spontaneous emission time and inhomogeneity must still be considered. This is also consistent with recent FWM experiments \cite{martin2018encapsulation}.

Finally, the control of the radiative lifetime resulting from the hBN encapsulation is further confirmed by measuring the dynamics of the charged exciton (trion, T). Figure~\ref{fig:fig3}(a) displays the normalized luminescence intensity dynamics of the charged exciton T for different hBN thicknesses in sample III. In contrast to the neutral exciton the variation of the bottom hBN thickness has here an impact on the trion luminescence decay time (and not on its rise-time). As the charged exciton oscillator strength is smaller than the neutral exciton one \cite{wang2014valley,chang2018crossover}, the trion radiative lifetime of the order of $\sim100$ ps is now longer than the relaxation/formation time. As a result, the PL rise time corresponding to this energy relaxation time does not vary with the cavity thickness. Here, the striking feature is that we find a variation of the trion PL decay time as a function of the hBN thickness very similar to the variation of the  neutral exciton radiative time, Fig.~\ref{fig:fig2}(b). Nevertheless, the amplitude of the variation is much smaller for the trion (typically ~10\%) whereas in the same sample the measured neutral exciton lifetime varies by more a factor two ($\sim 3$ to $7$ ps), Fig.~\ref{fig:fig2}(b).

The cavity effects revealed in this work make it possible to elucidate the complex dynamics of relaxation and recombination of excitons in TMD MLs \cite{brem2018exciton}. In general, the exciton lifetime $\tau$, measured in time-resolved luminescence dynamics, depends on both radiative and non-radiative (NR) recombination channels with $1/\tau =1/\tau_{rad} + 1/\tau_{nr}$. The radiative decay channel depends on the electrodynamical environment characteristics due to the Purcell effect while the non-radiative one, having no electromagnetic origin is assumed unchanged. Remarkably, the strong variation of the neutral exciton lifetime reported in Fig.~\ref{fig:fig2} demonstrates that the neutral exciton lifetime at low temperatures is limited by the radiative recombination (controlled here by the Purcell effect) with negligible contribution of NR channels.  However we did not observe any effect of the environment on the exciton dynamics for lattice temperatures above 80 K (see SI~\cite{suppl1}). This is due to the fact that the exciton lifetime is no more controlled by purely radiative recombination ~\cite{Selig:2016aa}. The rather small modulation of the trion lifetime observed in Fig.~\ref{fig:fig3} reveals that it is significantly affected by NR recombination. We can infer a NR trion recombination time of the order of $\tau_{nr}\sim100$ ps, i.e. competitive with the radiative one. 

Excellent fits of both the neutral and charged exciton PL dynamics can be obtained with the two-level model using the same relaxation time $\tau_{relax}$ from the photogenerated high energy states, inset of Fig.~\ref{fig:fig3}(b). As already reported for non-encapsulated TMD MLs~\cite{Robert:2016a}, we do not find here any evidence of electronic transfer from neutral excitons to trions in MoSe$_2$ ML. This result seems counterintuitive since the PL decay time of the neutral exciton X coincides with the measured PL rise time of the charged exciton, see Fig.~\ref{fig:fig3}(b), as if the X lifetime would be controlled by the trion formation time.  This behavior is simply due to the fact that the same energy relaxation time $\tau_{relax}$ drives both the neutral exciton PL decay time and charged exciton PL rise time (see SI~\cite{suppl1} where also alternative scenarios are discussed). 

In conclusion, we have shown that encapsulation of TMD MLs with hBN does not only improve the exciton emission/absorption linewidth by reducing the disorder-induced broadening related to local dielectric fluctuations. The hBN layers surrounding the semiconducting monolayer also have a dramatic impact on the exciton photon coupling through the Purcell effect. We demonstrate that we can control the radiative recombination time by one order of magnitude from $\sim 1$~ps up to about $10$~ps in full agreement with the theoretical analysis. This opens the way to engineer the exciton-photon coupling in these van der Waals heterostructures. An interesting prospect would be to deposit TMD monolayers on top of epsilon-near-zero metamaterials~\cite{Maas:2013aa} to obtain stronger enhancement of the exciton radiative decay rate. 

\emph{Acknowledgements.}   (*) HHF and BH have contributed equally to this work. We thank M. Gurioli, S. Berciaud and A. Poddubny for stimulating discussions. We acknowledge funding from ANR 2D-vdW-Spin, ANR VallEx, ANR MagicValley, Labex NEXT projects VWspin and MILO, ITN Spin-NANO No 676108 and ITN 4PHOTON Nr. 721394. M.A.S. and M.M.G. acknowledge partial support from LIA ILNACS through the RFBR project 17-52-16020. M.A.S. also acknowledges partial support of the Government of the Russian Federation (Project No. 14.W03.31.0011 at the Ioffe Institute). Growth of hexagonal boron nitride crystals was supported by the Elemental Strategy Initiative conducted by MEXT, Japan, and CREST (JP- MJCR15F3), JST. X.M. also acknowledges the Institut Universitaire de France.


\end{document}


\title{
Online supplemental materials: \\
Control of the Exciton Radiative Lifetime in van der Waals Heterostructures 
}

\author{H.H. Fang$^1$}
\author{B. Han$^1$}
\author{C. Robert$^1$}
\author{M.A. Semina$^2$}
\author{D. Lagarde$^1$}
\author{E. Courtade$^1$}
\author{T.~Taniguchi$^3$}
\author{K. Watanabe$^3$}
\author{T. Amand$^1$}
\author{B. Urbaszek$^1$}
\author{M.M. Glazov$^2$}
\author{X. Marie$^1$}

\affiliation{%
$^1$Universit\'e de Toulouse, INSA-CNRS-UPS, LPCNO, 135 Av. Rangueil, 31077 Toulouse, France}
\affiliation{$^2$Ioffe Institute, 194021 St. Petersburg, Russia}
\affiliation{$^3$National Institute for Materials Science, Tsukuba, Ibaraki 305-0044, Japan}
\maketitle

\tableofcontents

\section{Experimental Methods}

 \begin{figure}[t]
\includegraphics [width=\linewidth,keepaspectratio=true]{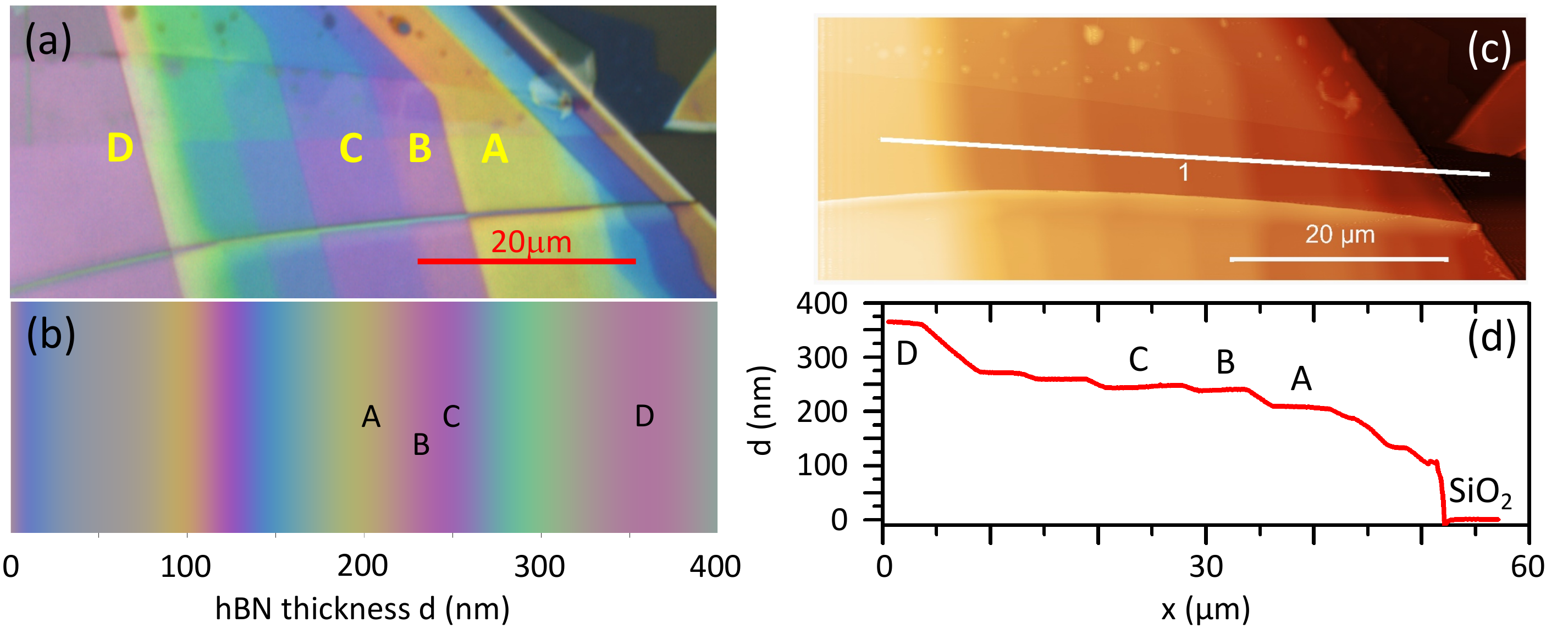}
\caption{(a) Optical microscope image of Sample III showing different terraces. (b) White light reflectivity: simulation of color of hBN on SiO$_2$(80nm)/Si with the transfer matrix method. (c-d). The thicknesses $d$ are $206$~nm~(A), $237$~nm~(B), $247$~nm~(C), $262$~nm~(D), $274$~nm~(E), and  $358$~nm~(F).}\label{fig:S1} 
\end{figure}

 \begin{figure}[h]
\includegraphics [width=\linewidth,keepaspectratio=true]{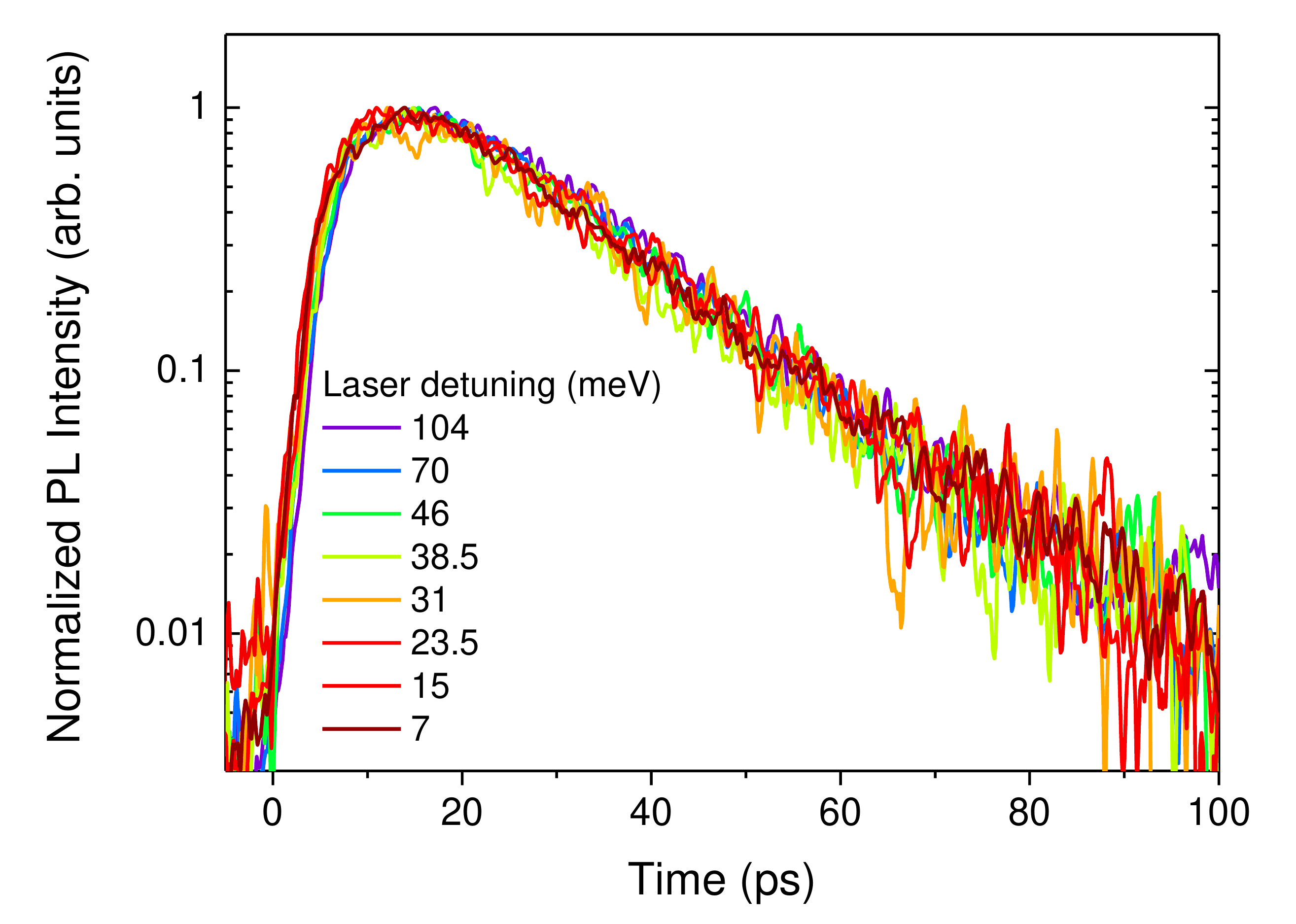}
\caption{Normalized exciton PL intensity dynamics for different laser excitation energies, i.e. different energy detuning with respect to the exciton resonance measured on a sample with $d=130 nm$ ($\tau_X=6$~ps).}\label{fig:S1:add} 
\end{figure}

We fabricate the van der Waals heterostructures by mechanical exfoliation of high quality hBN crystals  and bulk MoSe$_2$ (similar results have been obtained with commercial MoSe$_2$ provided by 2D Semiconductors or HQ Graphene). A first layer of hBN is mechanically exfoliated and deposited onto a 80 nm SiO$_2$/Si substrate using a dry-stamping technique. The deposition of the subsequent MoSe$_2$ ML and the second hBN capping layer is obtained by repeating this procedure. The in-plane size of the investigated MoSe$_2$ MLs is typically $\sim 10 \times 10~\mu$m$^2$. 
The samples are held on a cold finger at a temperature $T=7$ K  in a closed-cycle He cryostat. Attocube X-Y-Z piezo-motors allow for positioning with nm resolution of the ML with respect to the microscope objective (numerical aperture NA$=0.82$) used for the excitation and collection of luminescence. The \emph{cw} PL experiments are performed with a He-Ne laser (633 nm) for excitation focused onto a spot diameter of 1 $\mu$m. The PL signal is dispersed in a spectrometer and detected with a Si-CCD cooled camera. For time-resolved photoluminescence experiments presented in the main text, the flakes are excited by $\sim1.5$ ps pulses generated by a tunable mode-locked Ti:Sa laser with a repetition rate of 80 MHz and wavelength of 712 nm, i.e. about 100 meV above the neutral exciton transition energy. Similar results have been obtained for laser excitation wavelength in the range 710\ldots753 nm (laser detuning from exciton resonance ranging from 7 to 104 meV), see Fig.~\ref{fig:S1:add}. The PL signal is dispersed by a simple spectrometer and detected by a synchro-scan Hamamatsu streak camera~\cite{Lagarde:2014a,Robert:2016a}. By measuring the backscattered laser pulse from the sample surface, we obtain the overall instrumental response of the time-resolved setup, Fig. 2(a) in the main text.

\section{Measurement of the cavity thickness using AFM and reflectivity spectra}

The thicknesses of the bottom hBN layer $d$ is roughly determined by the color obtained in reflectivity Fig.~\ref{fig:S1}(a). The reflectivity spectrum on each terrace is fitted  using the calculated reflectivity obtained with the transfer matrix method, Fig.~\ref{fig:S1}(b).  AFM measurements confirm this preliminary determination and lead to more precise values, Fig.~\ref{fig:S1}(c) and (d).

\section{Bi-exponential fit of the time-resolved photoluminescence }

\begin{figure}[ht]
\includegraphics [width=\linewidth,keepaspectratio=true]{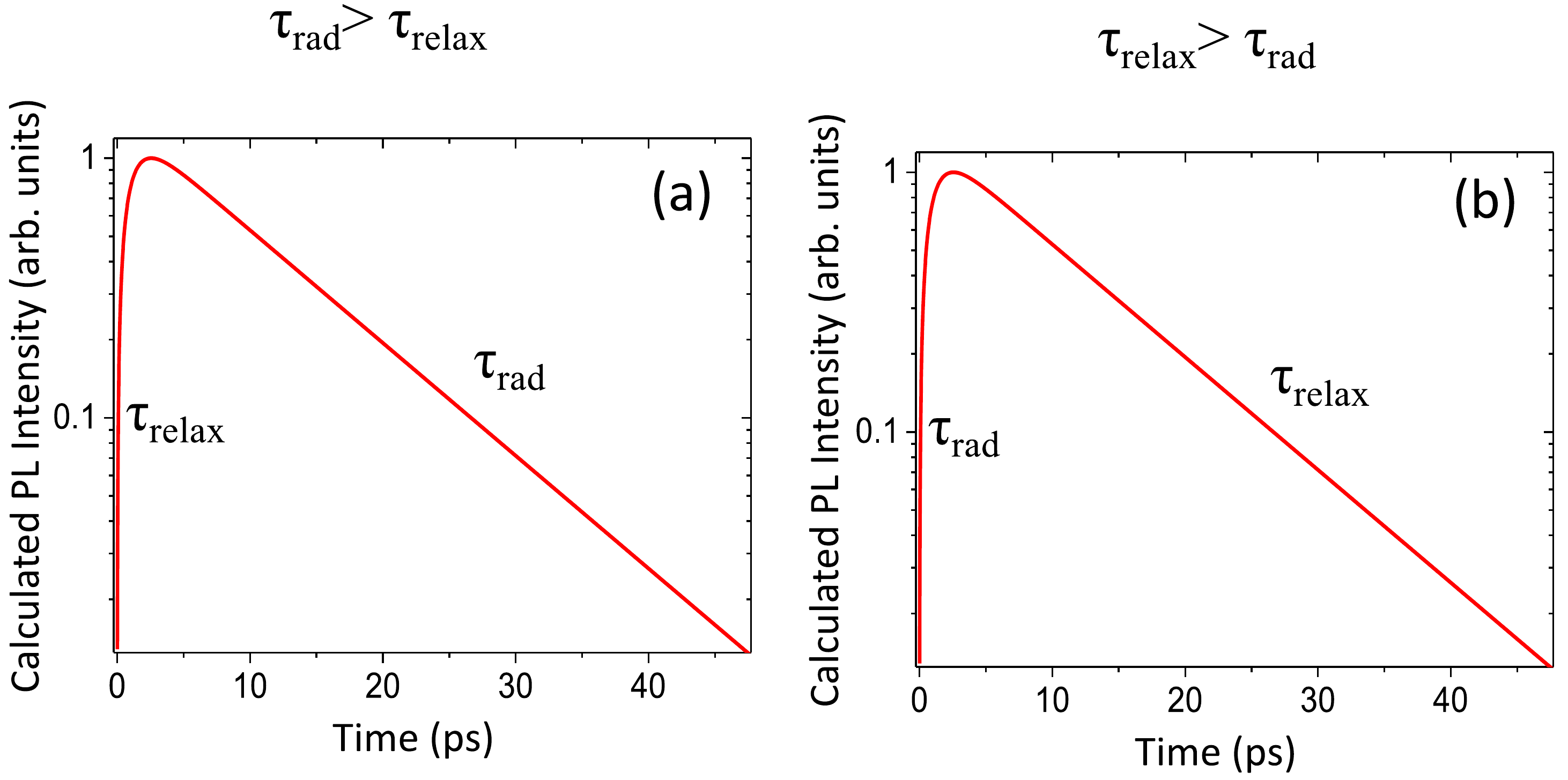}
\caption{Calculated PL intensity with (a) $\tau_{rad}>\tau_{relax}$ ($\tau_{rad}=10$~ ps, $\tau_{relax}=1$~ps) and (b) $\tau_{relax}>\tau_{rad}$ ($\tau_{rad}=1$~ps, $\tau_{relax}=10$~ps).  }\label{fig:S2} 
\end{figure}

The exciton kinetics are fitted with simple bi-exponential fits based on a simple two-level model assuming that the system can be described by the population of the ``reservoir'' of  photogenerated hot carriers $n_{hot}$ and of the cold excitons $n_{X}$ (inset of Fig. 3(b) in the main text):
\begin{subequations}
\label{biexp}
\begin{align}
&\frac{dn_{hot}}{dt}= -\frac{n_{hot}}{\tau_{relax}},\\
&\frac{dn_{X}}{dt}= -\frac{n_{X}}{\tau_{rad}}+\frac{n_{hot}}{\tau_{relax}},
\end{align}
\end{subequations}
where $\tau_{relax}$ is the relaxation rate of the hot excitons into the emitting states and $\tau_{rad}$ is the radiative decay rate of excitons. In Eqs.~\eqref{biexp} we disregarded non-radiative recombination processes. Then the calculated exciton PL intensity simply writes:
\begin{equation}
\label{Int}
I(t) \propto \frac{n_{hot}^{(0)}}{\tau_{relax}-\tau_{rad}} 
\left[\exp{\left(-\frac{t}{\tau_{relax}}\right)}-\exp{\left(-\frac{t}{\tau_{rad}}\right)} 
\right].
\end{equation}
Here $n_{hot}^{(0)}$ is the initial photogenerated population of the hot carriers. Note that depending on the relation between the relaxation and radiative times the rise time of PL and its decay time are controlled by different parameters. Figure~\ref{fig:S2} displays the corresponding calculated intensity for two cases: (a) $\tau_{rad}>\tau_{relax}$ and (b) $\tau_{relax}>\tau_{rad}$. The rise-time does correspond to the radiative recombination time if $\tau_{relax}>\tau_{rad}$.

 \begin{figure}[ht]
\includegraphics [width=\linewidth,keepaspectratio=true]{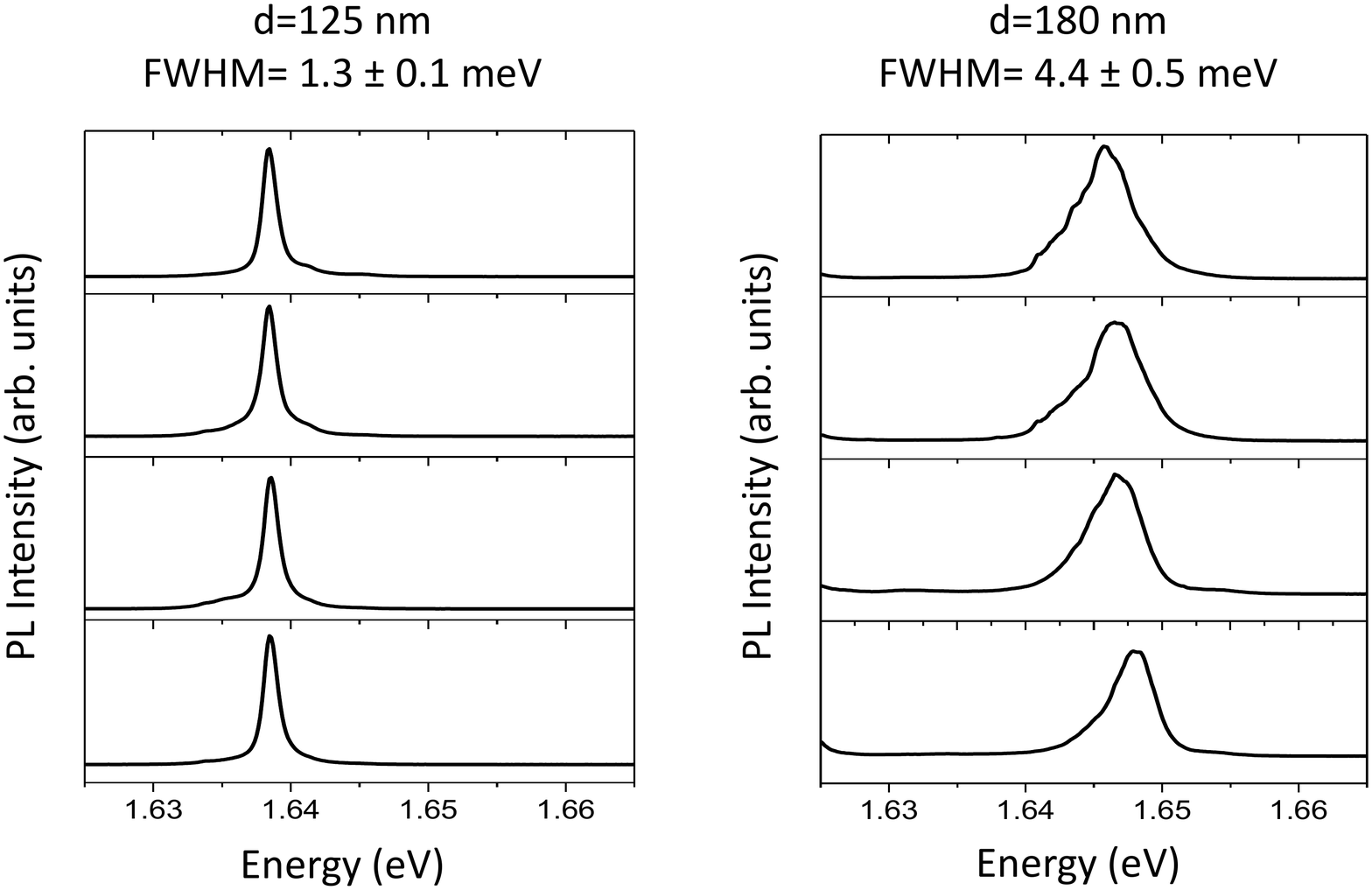}
\caption{ PL spectra at T=7 K on two terraces of sample IV corresponding to inhibition ($d=125$~nm) and enhancement ($d=180$~nm) of radiative decay rate. The linewidth measured on four different spots confirmed the trend of Fig.~2(c) of the main text showing narrower (larger) linewidth for monolayers located at node (anti-node) of the optical field.}\label{fig:S3} 
\end{figure}

\section{Cw photoluminescence spectra and estimation of the radiative lifetime in sample I}

The time-resolved photoluminescence measurements show that the radiative lifetime in sample I is limited by the temporal resolution (in the main text we infer $\tau_X < 1.5$~ps)
We can tentatively estimate the radiative lifetime by combining the \emph{cw} and time-resolved PL results.
The exciton homogeneous linewidth writes: $\Gamma=\Gamma_{rad} + \Gamma'$, where $\Gamma_{rad}$ is the radiative linewidth and $\Gamma'$ includes both non-radiative and pure-dephasing processes. The linewidth contributions from inhomogeneous broadening, light scattering, non-radiative processes, and radiative decay cannot be, unfortunately, disentangled by linear techniques such as photoluminescence or reflectivity spectroscopy used here. Nevertheless a rough estimate can be obtained assuming that the linewidth in sample II is mainly determined by pure dephasing processes  and neglecting inhomogeneous broadening. We find $\Gamma'\approx1.1$~meV (FWHM); the measured radiative lifetime ($\approx 10$ ps) gives a negligible radiative linewidth contribution : $\Gamma_{rad}\approx 0.06$~meV.  
Assuming that non-radiative and pure dephasing processes are identical in samples I and II, we can deduce a radiative linewidth of $\approx 0.9$ meV in sample I corresponding to a radiative lifetime of $\approx 740$~fs.
Although this analysis is rather approximate, the results are consistent with the recent FWM measurements in similar samples demonstrating that the homogenous broadening is not fully controlled by the radiative broadening \cite{martin2018encapsulation}, for example, disorder-induced broadening still plays a significant role.
The trend observed in Samples I and II is confirmed by the photoluminescence spectra measured in sample IV (same monolayer embedded in a cavity with different thicknesses), see Fig.~\ref{fig:S3}.

\section{Time-resolved photoluminescence of the neutral exciton at 90 K}

\begin{figure}[ht]
\includegraphics [width=\linewidth,keepaspectratio=true]{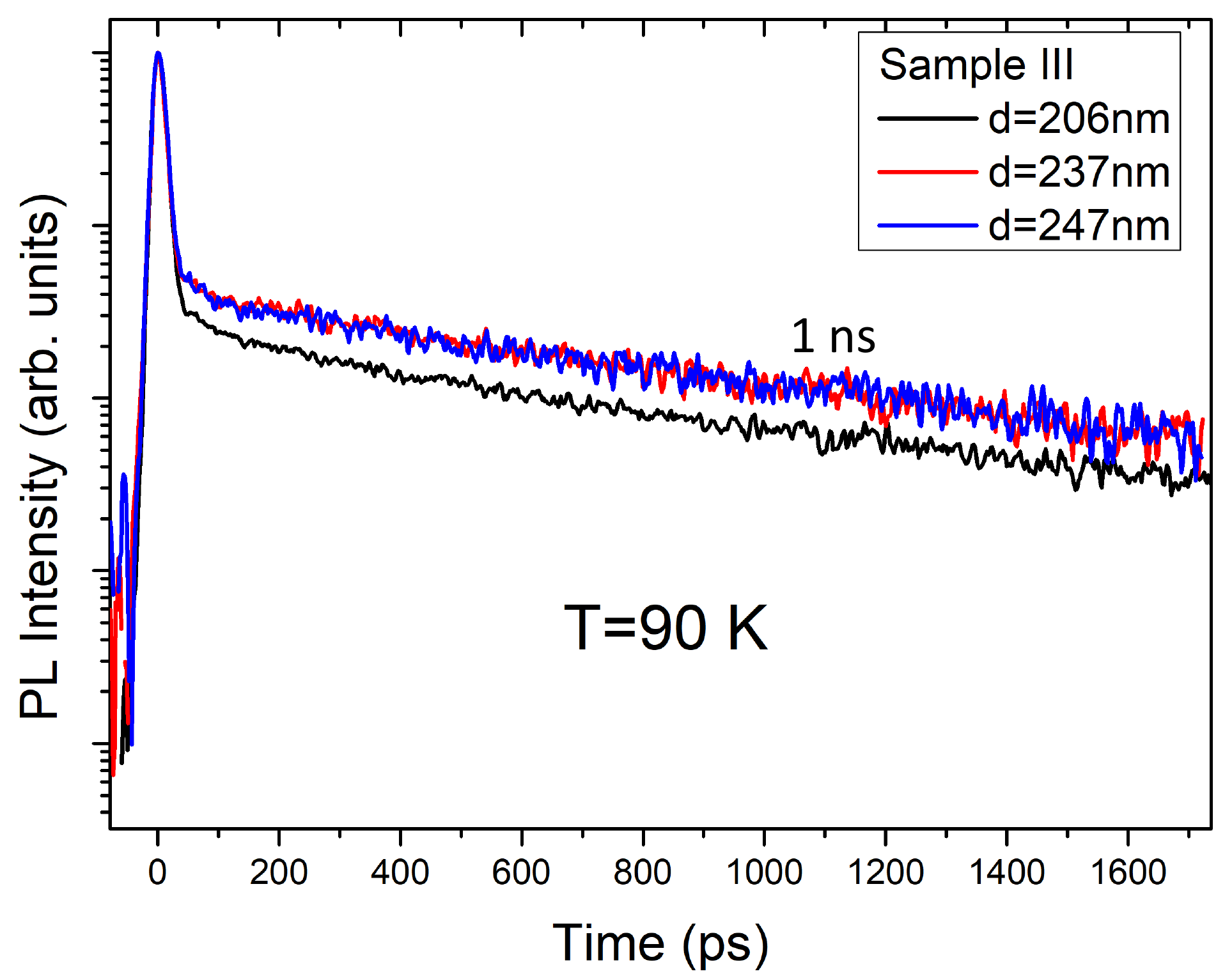}
\caption{Neutral exciton PL dynamics at T=90 K of sample III for three different bottom hBN thicknesses. At this temperature the decay time does not depend any more on the cavity thickness.}\label{fig:S4} 
\end{figure}

Figure~\ref{fig:S4} displays the neutral exciton PL dynamics at $T=90$~K of sample III for three different bottom hBN thicknesses $d$. Two regimes are observed. In the first one, the excitons are in a non-thermal regime and their dynamic is dominated by a competition between radiative rate and escape of the light cone through exciton-phonon interactions. The second regime corresponds to the long decay time of 1 ns related to the lifetime of thermalized excitons in agreement with  Ref.~\cite{Robert:2016a} . This decay time does not depend on the bottom hBN thickness showing that the decay rate of excitons at this temperature is no more dominated by radiative recombination, in contrast to the results presented in the main text at $T=7$~K.

 \begin{figure}[ht]
\includegraphics [width=0.9\linewidth,keepaspectratio=true]{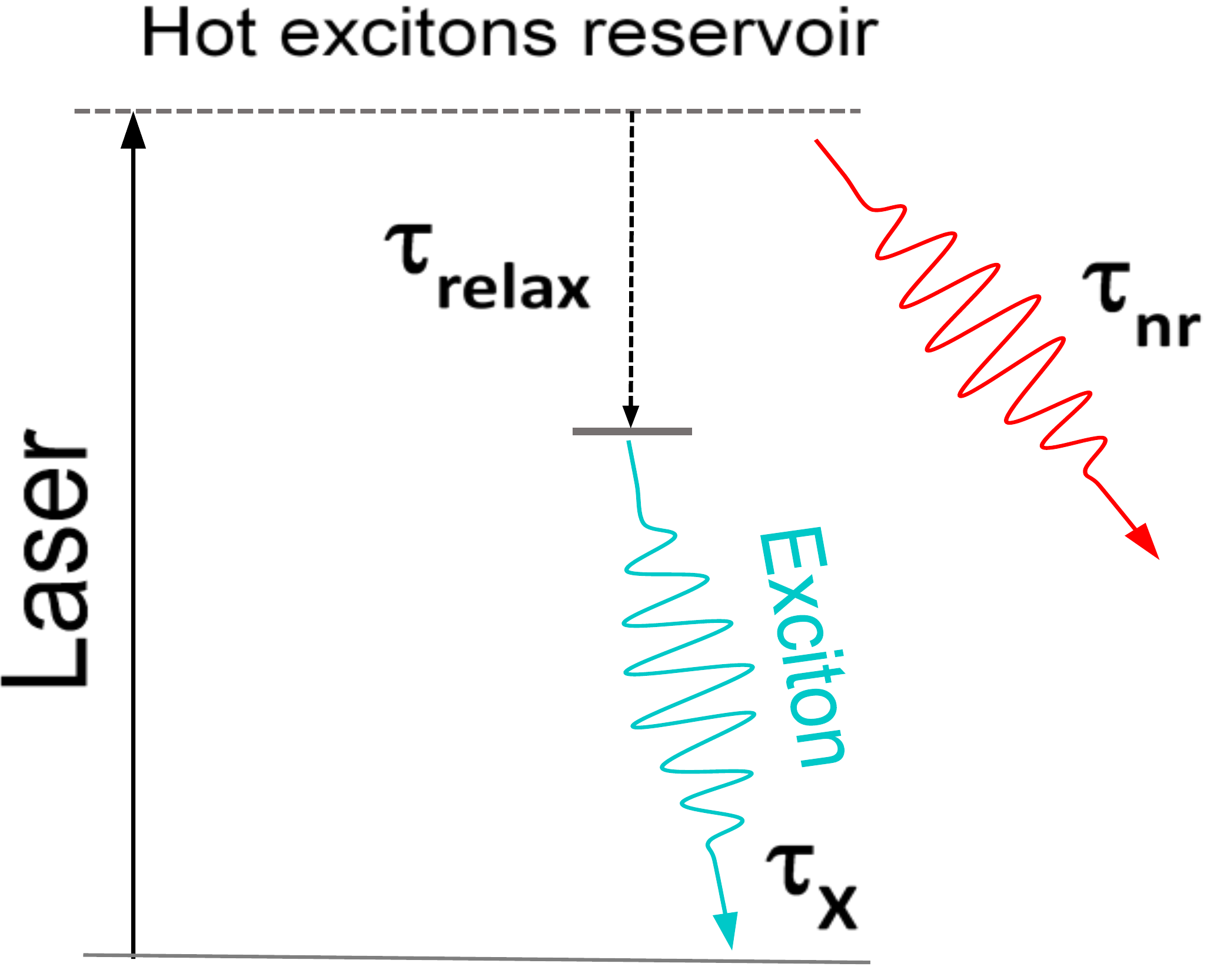}
\caption{Alternative scenario of the exciton dynamics which includes non-radiative processes.}\label{fig:S5:new} 
\end{figure}

\section{Possible role of non-radiative channels in the PL dynamics}

Our attribution of the decay time of neutral exciton to the relaxation of hot excitons in the main text assumes for simplicity the absence of non-radiative channels. Although this work demonstrates that the lifetime of an exciton in the radiative cone is dominated by its radiative recombination (tunable by cavity effects), we cannot claim that all hot excitons necessarily relax to the light cone. Indeed, any non-radiative channel such as defects trapping, relaxation to a dark state or formation of a trion at a timescale shorter than the relaxation to the light cone would lead exactly to the same dynamics but with smaller neutral exciton PL yield. In this subsection we briefly analyze this scenario depicted in Fig.~\ref{fig:S5:new} where these non-radiative decay processes of excitons are considered. The occupancies of the hot exciton reservoir and of the excitons in the light cone are given in this case by the following set of rate equations:
\begin{subequations}
\label{biexp:alternative}
\begin{align}
&\frac{dn_{hot}}{dt}= -\frac{n_{hot}}{\tau_{relax}} - \frac{n_{hot}}{\tau_{nr}},\\
&\frac{dn_{X}}{dt}= -\frac{n_{X}}{\tau_{rad}}+\frac{n_{hot}}{\tau_{relax}},\label{nx:cone}
\end{align}
\end{subequations}
Here the non-radiative decay rate of excitons includes contributions of trapping to defects, trion formation as well as possible formation of dark states.
For excitons within the light cone we neglect non-radiative processes since
\[
\tau_{nr} \gg \tau_{rad}.
\]
The calculation shows that the exciton PL dynamics under the condition 
\begin{equation}
\label{dec:alt}
\frac{1}{\tau_{decay}} = \frac{1}{\tau_{relax}}+\frac{1}{\tau_{nr}} \ll \frac{1}{\tau_{rad}}
\end{equation}
is given by the exponential law $\propto \exp{(-t/\tau_{decay})}$. We emphasize  that if the trion formation from the hot exciton reservoir is the dominant process, then it is not surprising to observe the same rise time for the trion than the decay time for the neutral exciton. 

\section{Reflection contrast}

\begin{figure}[ht]
\includegraphics [width=0.9\linewidth,keepaspectratio=true]{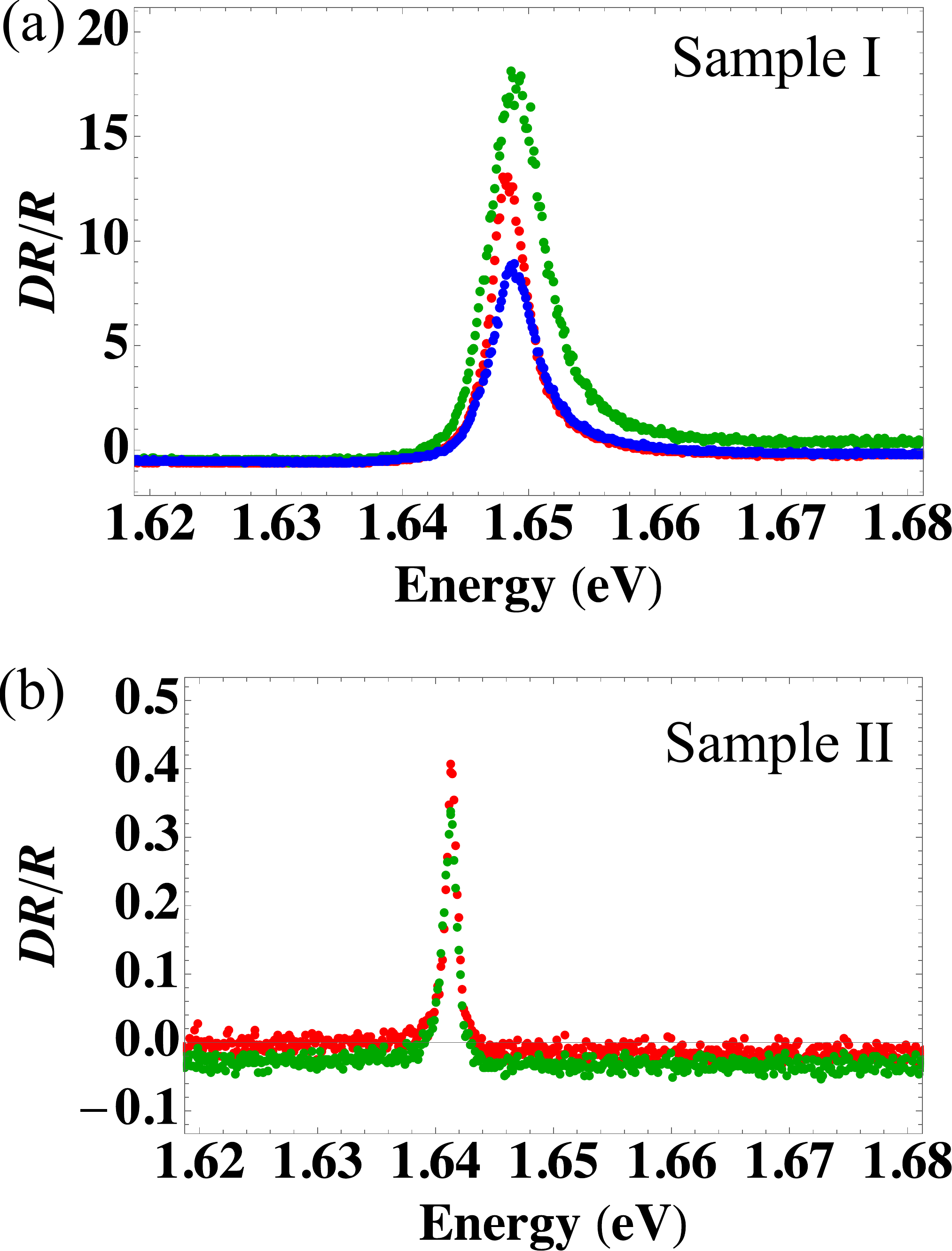}
\caption{Reflection contrast, $DR/R$, measured in (a) sample I  and (b) sample II  in the vicinity of the exciton resonance. Different sets of data points correspond to different spots on the sample surface.}\label{fig:S6:new} 
\end{figure}

In order to provide an additional evidence for the variation of the exciton radiative lifetime as a result of the cavity-like effect we present in Fig.~\ref{fig:S6:new} the measured reflectivity contrast 
\begin{equation}
\label{contrast:def}
\frac{DR}{R} = \frac{R_{ML} - R}{R},
\end{equation}
where $R_{ML}$ is the intensity reflection coefficient for the sample with the MoSe$_2$ monolayer and $R$ is the intensity reflection coefficient of the same sample but without the monolayer (measured in a different spot). Panel (a) shows the data on the sample I where the exciton radiative recombination is enhanced and panel (b) shows the data on the sample II where the exciton radiative recombination is inhibited. Despite certain spread of the $DR/R$ values, one can see that the reflection contrast in the sample I is systematically much larger than in the sample II ; the exciton linewidth in sample I is also substantially larger than the one in sample II. We stress that these reflectivity data are fully in line with the \emph{cw} PL and time-resolved PL measurements presented in the main text. We abstain from the detailed multi-parameter fit of the reflectivity data which requires also careful analysis of the light scattering and inhomogeneous broadening (as well as possible effects of finite NA of our optical setup).

\section{Theory of Purcell effect in van der Waals heterostructures}

In this section we outline the general electrodynamical method to calculate the radiative decay rate based on the transfer matrix formalism, present its justification on the basis of quantum mechanical approach.

\subsection{General approach using the transfer matrix method}\label{sec:gen}

We use the transfer matrix method in order to calculate the elementary response functions (reflectivity, transmission and absorbance) of our structure ``cap hBN layer/TMD ML/bottom hBN layer/SiO$_2$/Si''~\cite{robert2018optical}
\begin{multline}
\label{T:tot}
\hat T_{tot} = \hat T_{\rm SiO_2 \to Si} \hat T_{\rm SiO_2} \hat T_{\rm hBN \to SiO_2}  \hat T_{\rm hBN}'  \\
\hat T_{\rm MoSe_2} \hat T_{\rm hBN} \hat T_{air\to \rm hBN}.
\end{multline}
Here $\hat T_{i\to j}$ is the transfer matrix through the interface between the layers $i$ to $j$, $\hat T_i$ is the transfer matrix through the layer $i$, the prime denotes the bottom hBN layer. For the TMD ML transfer matrix we consider the situation of the monolayer embedded into the infinite hBN:
\begin{equation}
\hat T_{\rm MoSe_2} = \frac{1}{t}\begin{pmatrix}
{t^2-r^2} & r\\
-r & 1
\end{pmatrix},
\end{equation}
\begin{equation}
\label{rt}
r = \frac{\mathrm i \Gamma_0^{\rm hBN}}{\omega_0 - \omega - \mathrm i (\Gamma_0^{\rm hBN}+\Gamma)},\quad t = 1+r.
\end{equation}
Here $r$ and $t$ are the reflection and transmission coefficients of the TMD MLs in the infinite homogeneous hBN, $\omega_0$ is the exciton resonance frequency, $\Gamma_0^{\rm hBN}$ is the exciton radiative decay rate (into the hBN, $\Gamma_0^{\rm hBN} = \Gamma_0^{\rm vac}/n_{\rm hBN}$, where $n_{\rm hBN}$ is the hBN refractive index and $\Gamma_0^{\rm vac}$ is the radiative decay rate into free space, see Eq.~\eqref{ratio} and discussion below) and $\Gamma$ is the exciton non-radiative decay rate. The Si layer is assumed to be thicker than the absorption length, hence, the reflection of light at the interface Si and air is disregarded. The transfer matrix provides the following relation between  $r_{tot}$ and $t_{tot}$ the amplitude reflection and transmission coefficients through the structure:
\begin{equation}
\label{T:tot1}
\hat T_{tot}
\begin{pmatrix}
1\\
r_{tot}(\omega)
\end{pmatrix} = \begin{pmatrix}
t_{tot}(\omega)\\
0
\end{pmatrix}.
\end{equation}
Equation~\eqref{T:tot1} allows us to obtain $r_{tot}(\omega)$ and $t_{tot}(\omega)$ from the transfer matrix and the absorbance of the monolayer can be expressed as 
\begin{equation}
\label{Abs}
\mathcal A(\omega) = 1- |r_{tot}(\omega)|^2 - |t_{tot}(\omega)|^2/n_{\rm Si},
\end{equation}
where $n_{\rm Si}$ is the refraction index of Si.

According to the general theory~\cite{ll5_eng,ivchenko:2005a,Glazov:2014a} the poles of the response functions $r_{tot}(\omega)$, $t_{tot}(\omega)$ and, correspondingly, the feature in $\mathcal A(\omega)$, correspond to the eigenmodes of the system. In the vicinity of exciton resonance frequency the functions $r_{tot}(\omega)$ and $t_{tot}(\omega)$ can be recast in the form
\begin{subequations}
\label{poles}
\begin{align}
r_{tot}(\omega) = \frac{Z_r}{\omega_0^{\rm eff} - \omega -\mathrm i (\Gamma_0^{\rm eff}+ \Gamma)} + \ldots,\\
t_{tot}(\omega) = \frac{Z_t}{\omega_0^{\rm eff} - \omega -\mathrm i (\Gamma_0^{\rm eff}+ \Gamma)} + \ldots,
\end{align}
where dots denote regular part, $Z_r$ and $Z_t$ are the complex constants, and the quantities $\omega_0^{\rm eff}$ and $\Gamma_0^{\rm eff}$ represent the renormalized exciton frequency and its radiative decay rate. Accordingly, the absorbance in the vicinity of the resonance can be approximated by
\begin{equation}
\label{poles:a}
\mathcal A(\omega) \approx \frac{\mathcal A_0}{(\omega_0^{\rm eff} - \omega)^2+(\Gamma_0^{\rm eff}+ \Gamma)^2}.
\end{equation}
\end{subequations}
Thus, in order to calculate the radiative decay rate $\Gamma_0^{\rm eff}$  (HWHM)in our structure, we found numerically the absorbance, $\mathcal A(\omega)$ at $\Gamma\to 0$,\footnote{In this way we disregarded possible weak asymmetry of the optical spectra due to exciton-phonon interaction which could be taken into account by using $\Gamma$ as a function of frequency~\cite{Christiansen:2017a,shree2018exciton}.} fitted it by Eq.~\eqref{poles:a} and extracted $\Gamma_0^{\rm eff}$ as a function of the bottom hBN thickness $d$. These data are plotted by solid blue line in Fig.~2(b) of the main text.

\subsection{Fermi golden rule approach to emission into homogeneous medium}

Here and below we provide quantum mechanical approach to emission of excitons into homogeneous dielectric media. Let us start with the calculation of the emission rate of the exciton in a TMD ML into a free space. We neglect dielectric contrast between the ML and the surrounding. Hereafter we consider only $s$-polarization of light and fix the direction of the electric and magnetic fields.

Normalization of electric field per photon reads
\begin{equation}
\label{normE}
\frac{1}{8\pi} \int d\bm r \{|E_0|^2 + |B_0|^2 \} = \frac{ \mathcal V |E_0|^2}{4\pi} =  \frac{\hbar\omega}{2}, 
\end{equation}
\[
 |E_0| = |B_0| = \sqrt{\frac{2\pi \hbar\omega}{ \mathcal V}},
 \]
where $\mathcal V$ is the normalization volume, $E_0$ and $B_0$ are the complex amplidutes of the electric and magnetic field. It is convenient to use the vector potential $\bm A$ such that $\bm E = - c^{-1} \partial \bm A/\partial t$. Hence $\bm E_0 = (\mathrm i\omega /c) \bm A_0$ and the normalization for the vector potential reads
\begin{equation}
\label{normA}
 |A_0| = \sqrt{\frac{2\pi \hbar c^2}{\omega  \mathcal V}}.
\end{equation}
The light-matter coupling operator can be expressed as
\begin{equation}
\label{jA}
\hat V = - \frac{1}{c} \int d\bm r \bm j \bm A,
\end{equation}
where $\bm j$ is the current associated with exciton transition.
In the two-dimensional layer the matrix element of the perturbation~\eqref{jA} related to the formation of exciton per one photon with the center of mass wavevector $\bm K = (K_x, K_y) = 0$ reads 
 \begin{equation}
\label{Vexc}
\langle {\rm exc}| \hat V|0\rangle = - \sqrt{\mathcal S} \frac{e p_{cv}}{m_0 c} A_0 \varphi(0)\delta_{\bm q_\parallel,0}, 
\end{equation}
where $\mathcal S$ is the normalization area, $\varphi(0)$ is the exciton envelope wavefunction at the coinciding electron and hole coordinates, $p_{cv}$ is the interband momentum matrix element, $\bm q_\parallel = (q_x,q_y)$ is the in-plane wavevector of light. Using the Fermi golden rule we obtain the exciton radiative decay rate
\begin{equation}
\label{tau:0}
\frac{1}{\tau_0} = 2\Gamma_0^{\rm vac} = \frac{2\pi}{\hbar} \sum_{\bm q} |\langle {\rm exc}| \hat V|0\rangle|^2 \delta(\hbar c q - \hbar\omega_0).
\end{equation}
Here $\omega_0$ is the exciton resonance frequency, and the light wavevector $q=\sqrt{q_\parallel^2 + q_z^2}$ where $q_\parallel=0$ and $q_z \ne 0$ are the in-plane and normal components of the wavevector. We transform the sum over $\bm q$ into the integral over $q_z$ as
\begin{equation}
\label{transform}
\sum_{\bm q} \ldots \delta_{\bm q_\parallel,0} = \frac{\mathcal V}{\mathcal S} \int \frac{dq_z}{2\pi} \ldots \, .
\end{equation}
The energy conservation law provides two values of $q_z = \pm \omega_0/c$. As a result, for the radiative decay rate $\Gamma_0$ we have in agreement with Refs.~\cite{ivchenko:2005a,Glazov:2014a}
\begin{equation}
\label{Gamma0:vac}
\Gamma_0^{\rm vac} = \frac{2\pi q_0}{\hbar} \left(\frac{e p_{cv} \varphi(0)}{m_0 \omega_0} \right)^2.
\end{equation}
Here $q_0 = \omega_0/c$ is the wavevector of emitted radiation.

Let us now analyze the modifications of Eq.~\eqref{Gamma0:vac} for emission into a dielectric medium with the background dielectric constant $\varkappa_b$. Normalization of the field and vector potential reads (note that the relation between $\bm E$ and $\bm A$ does not depend on $\varkappa_b$ and in the homogeneous medium $B_0 = \sqrt{\varkappa_b} E_0$ so that electric and magnetic energies are the same):
\begin{equation}
\label{normEA}
\frac{\varkappa_b}{4\pi} \int d\bm r |E_0|^2  =  \frac{\hbar\omega}{2}, \quad |E_0| = \sqrt{\frac{2\pi \hbar\omega}{ \varkappa_b \mathcal V}}, \quad |A_0| = \sqrt{\frac{2\pi \hbar c^2}{ \varkappa_b \omega  \mathcal V}}.
\end{equation}
Additional modification in the Fermi golden rule approach comes from Eq.~\eqref{tau:0} where the energy conservation $\delta$-function reads now
\[
\delta(\hbar c q/\sqrt{\varkappa_b} - \hbar\omega_0) = \delta(\hbar c |q_z|/\sqrt{\varkappa_b} - \hbar\omega_0),
\]
therefore, the removal of the $\delta$-function yields $\sqrt{\varkappa_b}$ in the numerator. As a result~\cite{ivchenko:2005a,Glazov:2014a}:
\begin{equation}
\label{Gamma0:diel}
\Gamma_0(\varkappa_b) = \frac{2\pi q_b}{\varkappa_b\hbar} \left(\frac{e p_{cv} \varphi(0)}{m_0 \omega_0} \right)^2.
\end{equation}
Here $q_b = \omega_0 \sqrt{\varkappa_b}/c$ is the wavevector of emitted radiation in the medium. The comparison of Eqs.~\eqref{Gamma0:vac} and \eqref{Gamma0:diel} shows that 
\begin{equation}
\label{ratio}
\Gamma_0(\varkappa_b) = \frac{\Gamma_0^{\rm vac}}{\sqrt{\varkappa_b}}.
\end{equation}
In this approach we disregard, for simplicity, the modification of exciton wavefunction at the coinciding coordinates $\varphi(0)$, interband momentum matrix element $p_{cv}$ and the exciton resonance frequency $\omega_0$ due to the dielectric environment effects, including the screening of Coulomb potential of electron-hole interaction. This approach allows us to clarify the electrodynamical effects which can be interpreted here as: (i) change of the photon density of states and (ii) change of the field amplitude per photon.

\subsection{Comparison with electrodynamical approach}

\begin{figure}[t]
\includegraphics [width=\linewidth]{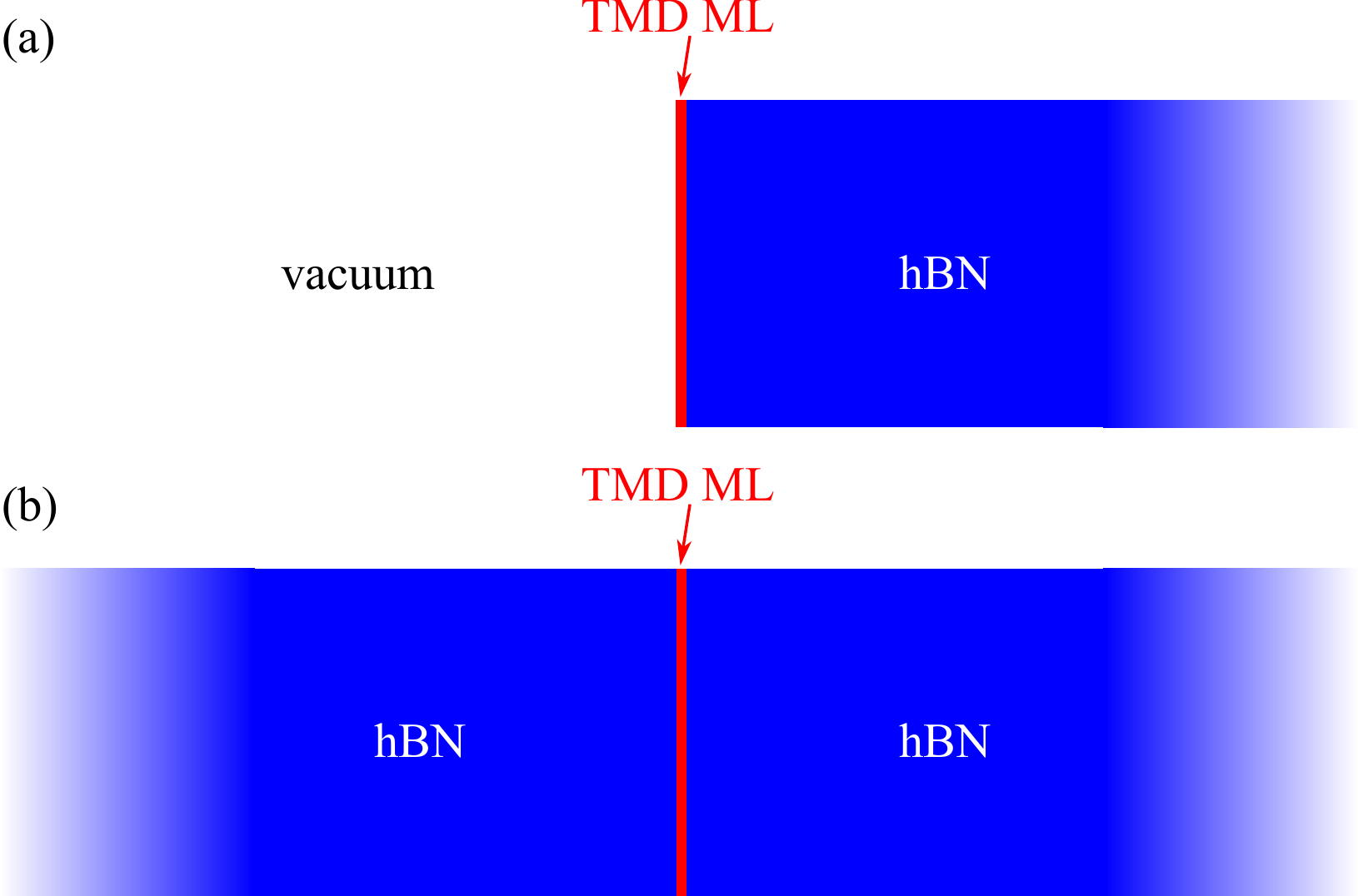}
\caption{(a) Semi-infinite structure consisting of a TMD ML on the hBN substrate. (b) TMD ML surrounded by the homogeneous hBN.}\label{fig:St1} 
\end{figure}

We demonstrate now that this result, Eq.~\eqref{ratio}, simply follows from the electrodynamical approach presented above. We consider a monolayer surrounded by the homogeneous hBN semi-infinite layers. We introduce
\begin{equation}
\label{bo}
r_{b0} = \frac{\sqrt{\varkappa_b}-1}{\sqrt{\varkappa_b}+1}, \quad r_{0b} = -r_{b0},
\end{equation}
the reflection coefficients from the dielectric to vacuum ($r_{b0}$) and from vacuum to the dielectric ($r_{0b}$). First, we calculate the reflection coefficient $r_+$ from the half space filled with a dielectric with the ML on its left side, Fig.~\ref{fig:St1}(a). Summing up all the reflections between the ML and the dielectric we arrive at 
\begin{equation}
\label{r+}
r_+ = r_0 + \frac{t_0^2 r_{0b}}{1-r_0 r_{0b}} = \frac{r_{0b}[\omega_0 - \omega - \mathrm i (-\Gamma_0^{\rm vac} + \Gamma)] + \mathrm i \Gamma_0^{\rm vac}}{\omega_0 - \omega - \mathrm i  [\Gamma_0^{\rm vac}(1+r_{0b})+\Gamma]},
\end{equation}
where $r_0$ and $t_0=1+r_0$ are the reflection transmission coefficient through the ML for light incident from the vacuum. These quantities differ from $r$ and $t$ in Eq.~\eqref{rt} by the replacement $\Gamma_0^{\rm hBN} \to \Gamma_0^{\rm vac}$ in Eq.~\eqref{Gamma0:vac}, see Ref.~\cite{ivchenko:2005a}. Next, summing up all the reflections between the left-hand side dielectric in Fig.~\ref{fig:St1}(b) and the structure ``ML+semi-infinite dielectric'' depicted in Fig.~\ref{fig:St1}(a) we arrive at 
\begin{equation}
\label{r:tot}
r_{tot} = r_{b0} + \frac{t_{b0}t_{0b} r_+}{1-r_{0b}r_+}.
\end{equation}
Here $t_{0b} = 1+r_{0b}$ and $t_{b0} = 1+r_{b0}$ are the corresponding transmission coefficients. Substitution of Eq.~\eqref{r+} into Eq.~\eqref{r:tot} gives
\begin{equation}
\label{r:tot1}
r_{tot} = \frac{\mathrm i \Gamma_0^{\rm vac}\frac{1+r_{0b}}{1-r_{0b}}}{\omega_0 - \omega - \mathrm i\left(\Gamma_0^{\rm vac}\frac{1+r_{0b}}{1-r_{0b}} +\Gamma\right)}.
\end{equation}
Equation~\eqref{r:tot1} has the same form as Eqs.~\eqref{rt} and provides the explicit connection between the renormalized and non-renormalized exciton radiative decay rate
\begin{equation}
\label{Gamma0:diel:ed}
\Gamma_0^{\rm hBN} = \Gamma_0^{\rm vac}\frac{1+r_{0b}}{1-r_{0b}} = \frac{\Gamma_0^{\rm vac}}{\sqrt{\varkappa_b}},
\end{equation}
in full agreement with Eq.~\eqref{ratio} confirming the electrodynamical approach.

\subsection{Emission in the semi-infinite dielectric structure}

Now we turn to the structure shown in Fig.~\ref{fig:St1}(a) where the semi-infinite dielectric is placed on the right of the ML. The reflection coefficient is given by Eq.~\eqref{r+}. In order to find the decay rate of exciton within the electodynamical approach it is sufficient to analyze the pole of this expression, cf. Eqs.~\eqref{poles}. Thus, for the radiative decay rate we have~\cite{Chance2007,khitrova:purcell}
\begin{equation}
\label{Gamma0:semi:ed}
\Gamma_0^{\rm semi} = \Gamma_0^{\rm vac} (1+\Re\{r_{0b}\}) = \Gamma_0^{\rm hBN}(1+\Re\{r_{b0}\}).
\end{equation}
This result was derived by the electrodynamical approach. Note, that this equality is general and can be applicable for any system with complex reflection coefficient, e.g., if the barrier is finite or the absorption is involved or if the barrier consists of several layers. In the particular case of a barrier made of a homogeneous dielectric with the real susceptibility $\varkappa_b$ we have
\begin{equation}
\label{Gamma0:semi:ed:11}
\Gamma_0^{\rm semi} = \Gamma_0^{\rm vac}  \frac{2}{1+\sqrt{\varkappa_b}} = \Gamma_0^{\rm hBN} \frac{2\sqrt{\varkappa_b}}{1+\sqrt{\varkappa_b}}.
\end{equation}

It is instructive to derive Eq.~\eqref{Gamma0:semi:ed} by means of the Fermi golden rule.  For convenience and in order to avoid problems with fact that the dispersion of waves in the left-hand side and in the right-hand side of the structure is different we assume that at $z\to + \infty$ the background dielectric constant $\varkappa_b$ goes smoothly to $1$. Asymptotic behavior of the modes of electromagnetic field at $z\to \pm \infty$ can be written as:
\begin{subequations}
\label{waves}
\begin{align}
E^{(1)}(z) = E_0 \times
\begin{cases}
1e^{\mathrm i q_z z} + r_{0b} e^{-\mathrm i q_z z}, \quad z\to -\infty \\
t_b e^{\mathrm i q_z z}, \quad z\to +\infty 
\end{cases}\\
E^{(2)}(z) = E_0 \times
\begin{cases}
t_b' e^{-\mathrm i q_z z}, \quad z\to -\infty \\
1e^{-\mathrm i q_z z} + r_{b}' e^{\mathrm i q_z z}, \quad z\to +\infty \\
\end{cases}
\end{align}
\end{subequations}
Here $|E_0|$ is given by Eq.~\eqref{normE},
the $t_b$ is the transmission coefficient through the dielectric barrier from the left to the right, $r_b'$ is the reflection coefficient from the barrier for the radiation incident from the right, $t_b'$ is the transmission coefficient through the barrier from the right to the left, see Fig.~\ref{fig:St1}(a).
Functions~\eqref{waves} are properly normalized which can be checked by means of the following relations~\cite{PhysRevA.43.2480}
\begin{equation}
\label{useful}
|r_{0b}|^2+|t_b|^2=1, \quad |r_{b}'|^2+|t_b'|^2=1, \quad r_{0b}^*t_b'+r_b't_b^*=0,
\end{equation}
which hold true for non-absorbing media only, where, strictly speaking, quantum mechanical approach is applicable. The dispersion relation remains the same as for the waves in the empty space.

Now we apply the Fermi golden rule [cf. Eq.~\eqref{tau:0}]
\begin{equation}
\label{tau:0:1}
\frac{1}{\tau_0^{\rm semi}} = 2\Gamma_0^{\rm semi} = \frac{2\pi}{\hbar} \sum_{q_z,i} |\langle {\rm exc}| \hat V^{(i)}|0\rangle|^2 \delta(\hbar c q_z - \hbar\omega_0),
\end{equation}
where $V^{(i)}$ is the perturbation for interaction with the waves~\eqref{waves}, $i=1,2$. The summation over $q_z$ is understood as follows [cf. Eq.~\eqref{transform}]
\begin{equation}
\label{transform:1}
\sum_{q_z,i} \ldots = \frac{\mathcal V}{\mathcal S} \sum_i \int_0^{\infty} \frac{dq_z}{2\pi} \ldots \, .
\end{equation}
Taking into account that the amplitude of the electromagnetic field at the TMD ML for the mode $1$ is given by $E_0(1+r_{0b})$ and for the wave $2$ is given by $E_0 t_b'$, the ratio
\begin{equation}
\label{Gamma0:semi:gen}
\frac{\Gamma_0^{\rm semi}}{\Gamma_0^{\rm vac}} = \frac{|1+r_{0b}|^2+|t_b'|^2}{2}.
\end{equation}
Equation~\eqref{Gamma0:semi:gen} can be recast in the form \eqref{Gamma0:semi:ed} making use of the following relations $|t_b'|^2 = |t_b|^2$, which follows from the time-reversal symmetry,  and Eq.~\eqref{useful}:
\begin{multline}
\label{main}
|1+r_{0b}|^2+|t_b'|^2 = |1+r_{0b}|^2+|t_b|^2 \\
= 1+|r_{0b}|^2 +2 \Re\{r_{0b}\}+|t_b|^2 = 2 +2 \Re\{r_{0b}\}.
\end{multline}
Thus, Fermi golden rule, Eq.~\eqref{Gamma0:semi:gen} and the electrodynamical approach, Eq.~\eqref{Gamma0:semi:ed}, gives the same result for non-absorbing dielectric media. It is worth stressing that Eq.~\eqref{Gamma0:semi:ed} is more general in a sense that it can be applied for absorbing structures as well where the quantum mechanical approach fails~\cite{khitrova:purcell}.

\section{Analytical model applied to our structure}

In this section we provide simple but accurate analytical approximation for the Purcell effect in our structure. We also consider the effects related to emission at oblique angles and demonstrate that they are not particularly important in the studied system.

\begin{figure}[ht]
\includegraphics [width=0.9\linewidth,keepaspectratio=true]{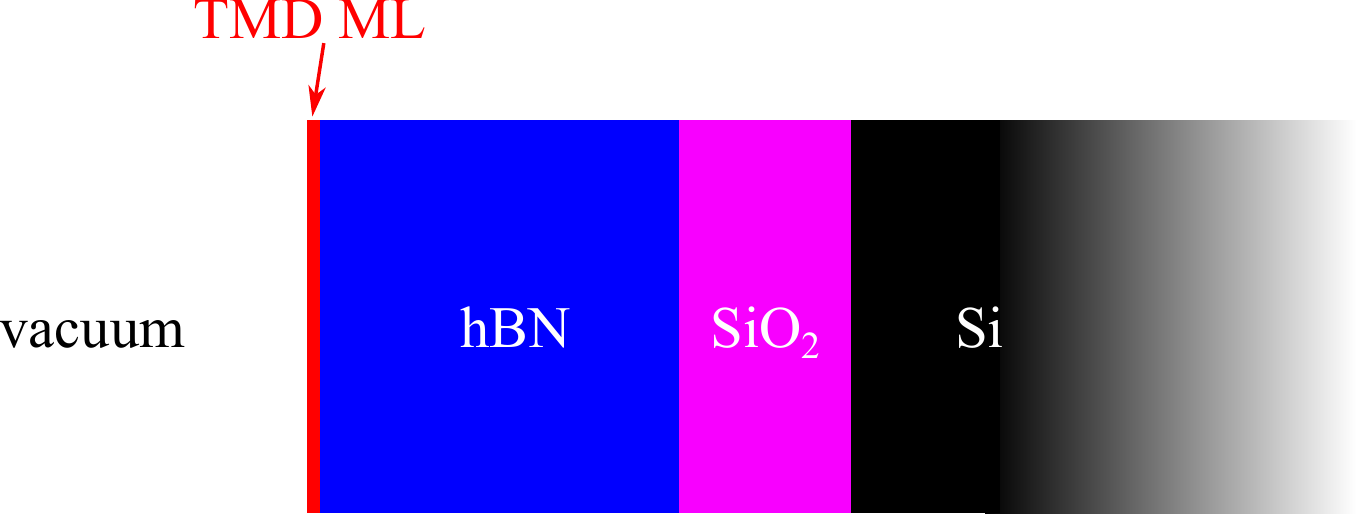}
\caption{Simplified structure used in our analytical model.}\label{fig:St2} 
\end{figure}

The main simplification comes from the fact that the thickness of the top hBN layer (typically within 10~nm range) does not affect electrodynamical properties of the structure because it is much smaller than the light wavelength at the exciton resonance frequency. That is why we can consider a simplified structure in our analytical model shown in Fig.~\ref{fig:St2}, which differs from the real sample [Fig. 1(a) of the main text] by the absence of the top hBN layer. Thus, in accordance with Eq.~\eqref{Gamma0:semi:ed} in order to find the exciton radiative decay rate we need to calculate the reflection coefficient $r_{bg}$ for the three-layer structure (bottom) hBN/SiO$_2$/Si for the light incident from the left.

\begin{figure}[h]
\includegraphics[width=0.8\linewidth]{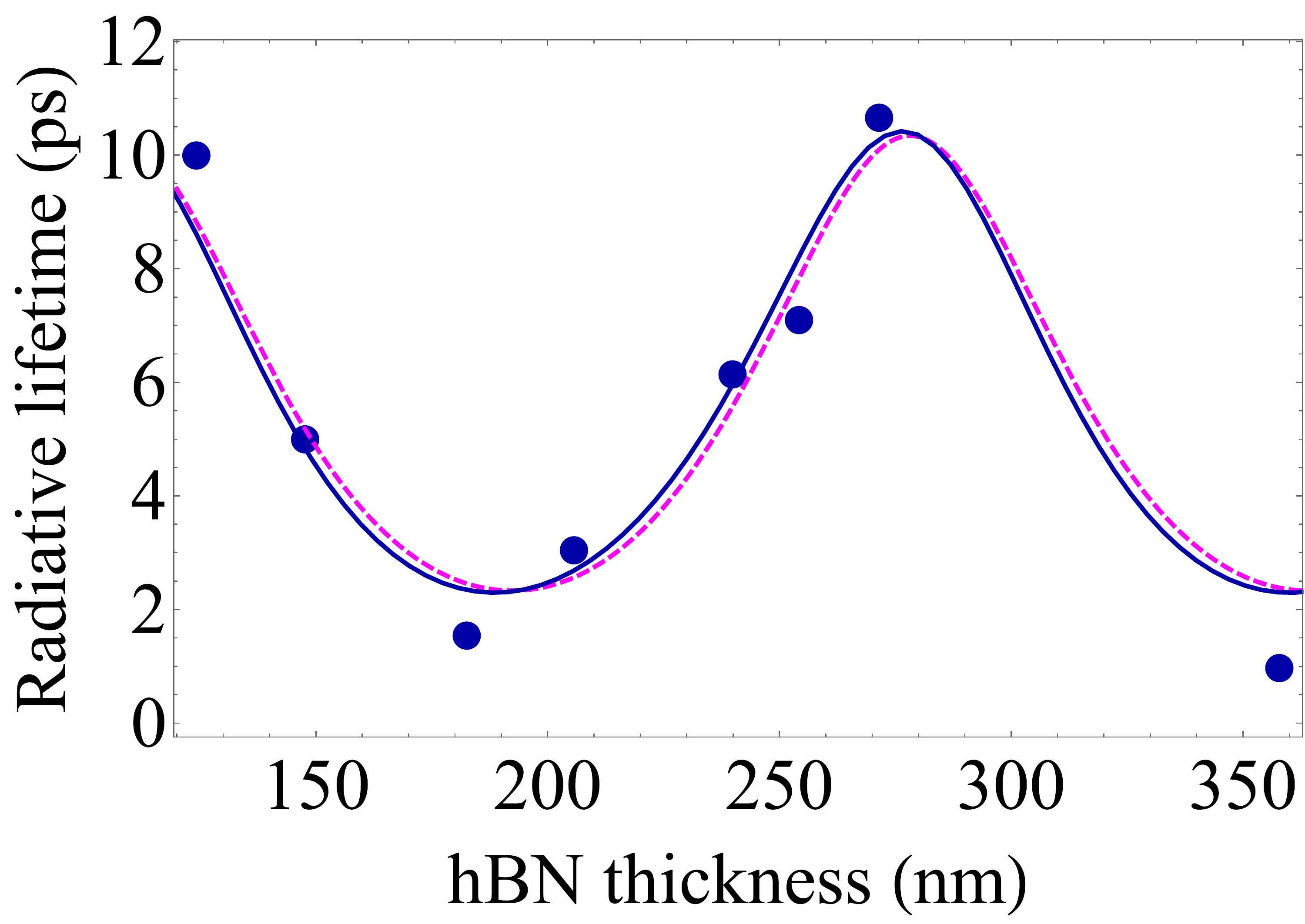}
\caption{Radiative lifetime as a function of hBN thickness. The parameters are: $1/(2\Gamma_0^{\rm vac}) = 2.7$~ps and SiO$_2$ thickness is 83~nm, $\hbar\omega_0 = 1.64$~meV. Dark blue solid line shows full calculation with the transfer matrix method (which includes also cap hBN layer of 5~nm, see Fig.~2(b) of the main text and Sec.~\ref{sec:gen}), magenta dashed line is the analytical calculation after Eqs.~\eqref{Fp} and Eq. (1) of the main text. Points are the experimental data, see Fig.~2(b) of the main text.}\label{fig:St3}
\end{figure}

We introduce the following notations for the reflection coefficients:
\begin{subequations}
\label{notations}
\begin{align}
&r_1 = \frac{1-n_{\rm hBN}}{1+n_{\rm hBN}}, \quad \mbox{vacuum $\to$ hBN},\\
&r_2 = \frac{n_{\rm hBN} - n_{\rm SiO_2}}{n_{\rm hBN} + n_{\rm SiO_2}}, \quad \mbox{hBN $\to$ SiO$_2$},\\
&r_3 = \frac{n_{\rm SiO_2} - n_{\rm Si}}{n_{\rm SiO_2} + n_{\rm Si}}, \quad \mbox{SiO$_2$ $\to$ Si},
\end{align}
\end{subequations}
we use $r_j' = -r_j$ for the reflection through the same interface but in the backwards direction, and we use $t_j=1+r_j$, $t_j' = 1+r_j'$ for the transmission coefficients, $j=1,\ldots, 3$. We use $q_j$ and $L_j$ to denote corresponding wavevector and thickness, i.e., $q_1$ is the wavevector in hBN, $L_1\equiv d$ (in the notations of the main text) is its thickness. For the SiO$_2$/Si part of the structure (light is incident from hBN) the reflection coefficient reads
\begin{equation}
\label{r23}
r_{23} = r_2 + \frac{t_2t_2'r_3 e^{2\mathrm i q_2L_2}}{1-e^{2\mathrm i q_2L_2}r_3r_2'}.
\end{equation}
Analogously, for the whole structure hBN/SiO$_2$/Si
\begin{equation}
\label{r123}
r_{bg} = r_1 + \frac{t_1t_1'r_{23} e^{2\mathrm i q_1L_1}}{1-e^{2\mathrm i q_1L_1}r_{23}r_1'}.
\end{equation}
The Purcell factor, i.e., the ratio of the emission rate in our structure and in vacuum, in agreement with written above is given by
\begin{equation}
\label{Fp}
F_p = \frac{\Gamma_0^{\rm eff}}{\Gamma_0^{\rm vac}}  = 1 + \Re\{r_{bg}\}.
\end{equation}
We stress that we analyze here the impact of the electrodynamical environment only. The effects related to possible variation of the exciton binding energy and oscillator strength due to modification of the Coulomb interaction dielectric screening are disregarded.
The calculation of the exciton radiative lifetime after Eq.~\eqref{Fp} and Eq.~(1) of the main text is presented in Fig.~\ref{fig:St3} together with the experimental data and the full numerical model. The difference between the full model and the analytical approximation is almost negligible and results from the non-zero thickness of the top hBN layer.

\subsection{Effects of the in-plane wavevector of exciton}

The aim of this section is to determine the influence of the emission angle because in the experiments, a high NA objective is used which collects photons emitted by excitons with non-zero momentum. To that end, let us now consider the emission of exciton with an in-plane wavevector $\bm k$. In this case, the polarization of radiation becomes important. As before, we start from the emission into free space and present the side-by-side comparison of the Fermi golden rule approach and electrodynamical treatment of the problem. The light-exciton coupling matrix elements depend on the light polarization [cf. Eq.~\eqref{Vexc}] and read:
\begin{subequations}
\label{V0exc}
\begin{align}
&s-\mbox{pol.:} \quad \langle {\rm exc}| \hat V_s|0\rangle = - \sqrt{\mathcal S} \frac{e p_{cv}}{m_0 c} A_0 \varphi(0)\delta_{\bm q_\parallel,\bm k}, \\
&p-\mbox{pol.:} \quad \langle {\rm exc}| \hat V_p|0\rangle = - \cos{\vartheta}\sqrt{\mathcal S} \frac{e p_{cv}}{m_0 c} A_0 \varphi(0)\delta_{\bm q_\parallel,\bm k},
\end{align}
\end{subequations}
where $\cos{\vartheta} = q_z/q = \sqrt{k^2 - q^2}/q$. Additionally, in the Fermi golden rule the energy conservation $\delta$-function should be transformed as
\begin{multline}
\label{delta:oblique}
\delta(\hbar c \sqrt{q_z^2+q_\parallel^2} - \hbar\omega_0) = \\
\frac{1}{\hbar c} \frac{\sqrt{q_z^2+q_\parallel^2}}{|q_z|} \delta\left(|q_z|-\frac{1}{c} \sqrt{\omega_0^2 - c^2 k^2}\right),
\end{multline}
which produces extra factor $1/\cos{\vartheta}$ in the Fermi golden rule. This factor arises from van Hove-like singularity of photon density of states at the edge of the light cone.
As a result, we have the following relations
\begin{equation}
\label{Gamma0:sp}
\Gamma_{0,s}^{\rm vac}(k) = \frac{1}{\cos{\vartheta}} \Gamma_0^{\rm vac}, \quad \Gamma_{0,p}^{\rm vac}(k) = {\cos{\vartheta}} \Gamma_0^{\rm vac}.
\end{equation}
This result can be also obtained from the electrodynamical approach~\cite{ivchenko:2005a,Glazov:2014a}. Similar relations (with the replacement $\rm vac\to \rm hBN$) hold for emission in the homogeneous hBN. The enhancement of $\Gamma_{0,s}$ for the grazing incidence ($|\vartheta|\to \pi/2$, $\cos{\vartheta} \to 0$) results from the simple fact that for light propagating in the ML plane the light-matter length in $s$-polarization becomes infinite.

\begin{figure*}[t]
\includegraphics[width=0.4\linewidth]{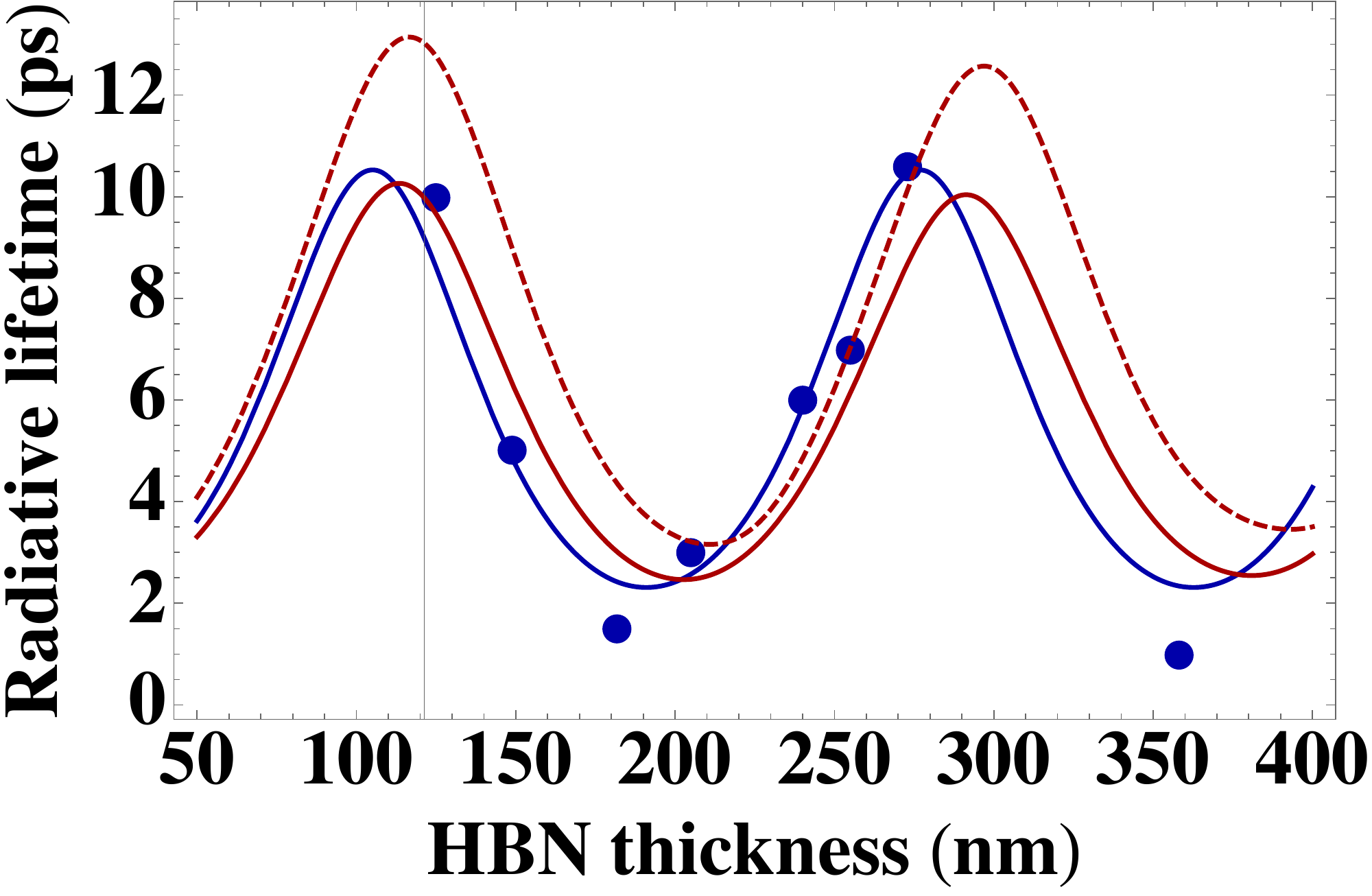}
\includegraphics[width=0.4\linewidth]{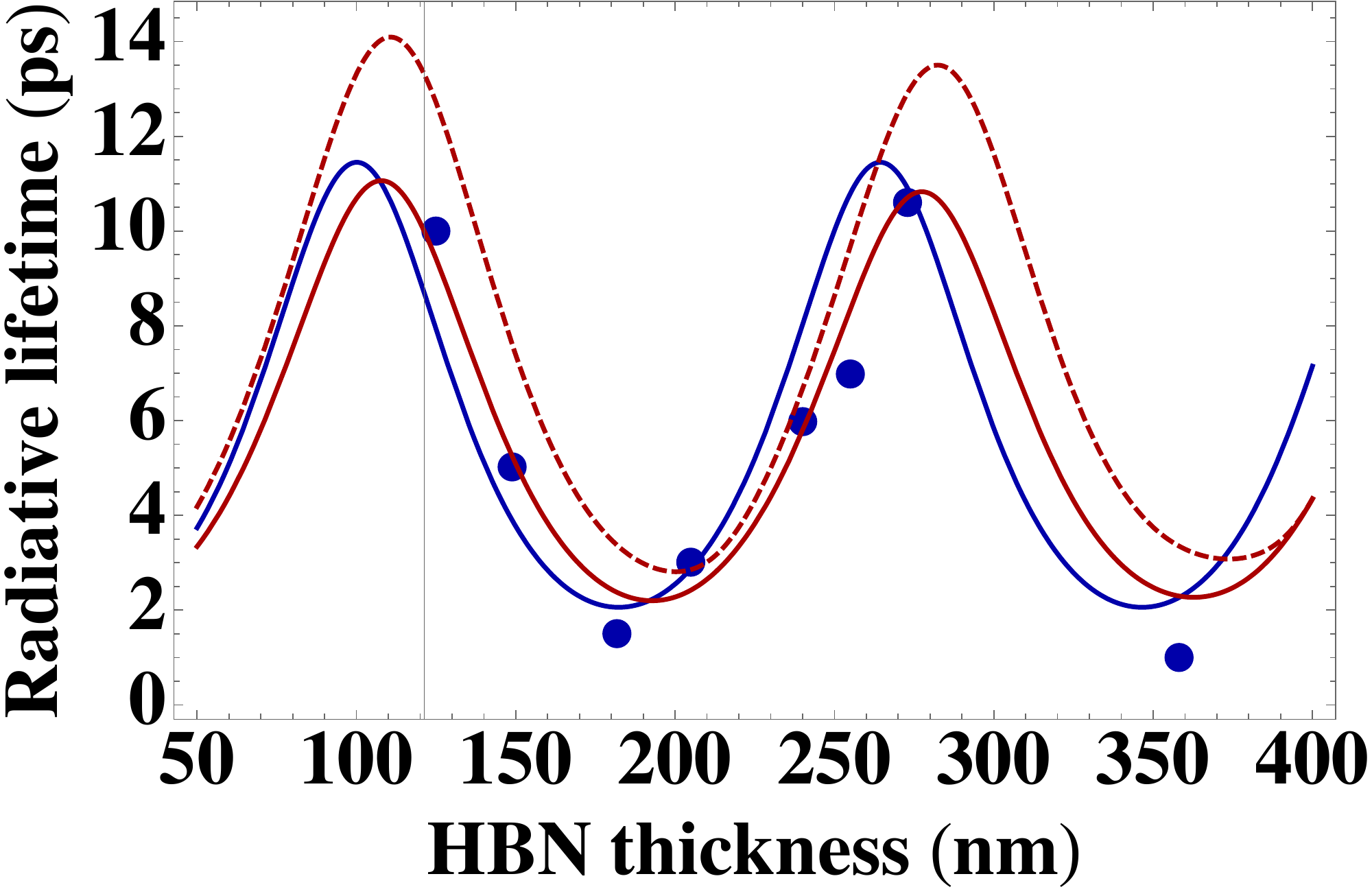}
\caption{Radiative lifetime as a function of hBN thickness calculated after Eq.~\eqref{Gamma:aver:bg:thetac} with the following parameters SiO$_2$ thickness is 83~nm, $\hbar\omega_0 = 1.64$~meV (same for all curves). Blue curve corresponds to collection angle $\vartheta_c =0$ (emission along the normal), solid brown corresponds to $\vartheta_c = 55^\circ$ (as in the experiment) and dashed brown corresponds to $\vartheta_c = 90^\circ$. Left panel corresponds to $1/(2\Gamma_0^{\rm vac}) = 2.7$~ps and previously used varied background refractive indices: $n_{\rm hBN} = 2.2$, $n_{\rm SiO_2}=1.46$, $n_{\rm Si} = 3.5$. Right panel corresponds to $1/(2\Gamma_0^{\rm vac}) = 2.5$~ps and slightly varied background refractive indices: $n_{\rm hBN} = 2.3$, $n_{\rm Si} = 3.9$. Points are the experimental data, see Fig.~2(b) of the main text. }\label{fig:St4}
\end{figure*}

The average decay rate for exciton population characterized by the distribution function $f(k)$ can be written as:
\begin{equation}
\label{Gamma:aver}
\bar\Gamma_0^{\rm vac} = \frac{1}{2 \sum_{|\bm k|\leqslant \omega/c} f(k)}\sum_{|\bm k|\leqslant \omega/c} f(k)\left[\Gamma_{0,s}^{\rm vac}(k)+\Gamma_{0,p}^{\rm vac}(k)\right]. 
\end{equation}
In the averaging we take into account the population of states within the light cone and assume that the distribution of exciton is independent of their polarization.
If $f(k)$ is constant within the light cone, which is reasonable due to the fact that the light cone is very ``narrow'' in the energy space, the integrals can be readily calculated via the substitution $\cos{\vartheta} = \mu$ with the result 
\begin{equation}
\label{Gamma:aver:1}
\bar\Gamma^{\rm vac}_0 = \frac{\Gamma_0^{\rm vac}}{2} \int_0^1 d\mu \mu\left(\mu^{-1} + \mu\right) = \frac{4}{3} \Gamma_0^{\rm vac}.
\end{equation}

Equation~\eqref{Gamma:aver} can be easily extended for our system depicted in Fig.~\ref{fig:St2}. To that end we introduce $r_{bg}^{(s)}$ and $r_{bg}^{(p)}$ for the reflection coefficients of our structure in $s$- and $p$-polarizations, respectively. It follows from Eqs.~\eqref{Fp} and~\eqref{Gamma:aver} that
\begin{widetext}
 \begin{equation}
\label{Gamma:aver:bg}
\bar \Gamma_0^{\rm eff} = \frac{n_{\rm hBN}}{2 \sum_{|\bm k|\leqslant \omega/c} f(k)}
\sum_{|\bm k|\leqslant \omega/c} f(k)\left[\Gamma_{0,s}^{\rm hBN}(k)\left(1+\Re\{r_{bg}^{(s)}\}\right)+\Gamma_{0,p}^{\rm hBN}(k)\left(1+\Re\{r_{bg}^{(p)}\}\right)\right]. 
\end{equation}
\end{widetext}

For completeness we present the expressions for reflection coefficients in $s-$ and $p-$polarizations at the boundaries of homogeneous dielectrics. We use the following notations for the wavevectors
\begin{equation}
\label{not:wavevectors}
q_j = \frac{\omega}{c} n_j, \quad  q_{z,j} = \sqrt{q_j^2 - k^2}, \quad j=0,1,2,3,
\end{equation}
$j=0$ corresponds to vacuum, $n_0=1$, $j=1$ is the hBN, $n_1 = n_{\rm hBN}$ (note that we disregard optical anisotropy of hBN, which can lead to quantitative changes), $n_2 = n_{\rm SiO_2}$, and $n_3 = n_{\rm Si}$. We present the reflection coefficients for light incident from the layer $j$ to $j+1$ as
\begin{equation}
\label{rsp}
r_{j,j+1}^{(s)} = \frac{q_{z,j} - q_{z,j+1}}{q_{z,j} + q_{z,j+1}}, \quad r_{j,j+1}^{(p)} = - \frac{n_{j+1}^2 q_{z,j} - n_{j}^2 q_{z,j+1}}{n_{j+1}^2 q_{z,j} + n_{j}^2 q_{z,j+1}}.
\end{equation}
The transmission coefficients read $t_{j,j+1}^{(s)}=1+r_{j,j+1}^{(s)}$ and $t_{j,j+1}^{(p)}=1+r_{j,j+1}^{(p)}$. Note that we use extra minus sign in the definition of $r_p$ in order to have $r_s = r_p$ at the normal incidence.

We also note that in actual calculations is it convenient to change from the integration over $\bm k$ to the integration over the angle of incidence $\vartheta$. Assuming that $f(k)$ is constant in the relevant wavevector range, by virtue of $k dk  = q_0^2 \sin{\vartheta} d\sin{\vartheta}  = q_0^2  \cos{\vartheta}\sin{\vartheta} d\vartheta$ the integral in Eq.~\eqref{Gamma:aver:bg} can be recast as ($\mu = \cos{\vartheta}$):
 \begin{equation}
\label{Gamma:aver:bg:1}
F_p = \int_0^{1} d\mu \left[\left(1+\Re\{r_{bg}^{(s)}\}\right)+\mu^2 \left(1+\Re\{r_{bg}^{(p)}\}\right)\right]. 
\end{equation}

In experiments a finite collection angle is used, $\vartheta_c<\pi/2$. In order to describe such a situation we suggest to use corresponding partial averaging and obtain the Purcell factor in the form
\begin{equation}
\label{Gamma:aver:bg:thetac}
F_p(\vartheta_c) = \frac{3}{4-3\cos{\vartheta}-\cos^3{\vartheta_c}} \times 
\end{equation}
\[
\int_0^{\arccos{\vartheta_c}} d\mu \left[\left(1+\Re\{r_{bg}^{(s)}\}\right)+\mu^2 \left(1+\Re\{r_{bg}^{(p)}\}\right)\right],
\]
where the prefactor comes from the normalization condition for emission into a homogeneous environment:
\[
\int_0^{\vartheta_c} d\vartheta_c \sin{\vartheta} (1+\cos^2{\vartheta}) = \frac{4-3\cos{\vartheta_c}-\cos^3{\vartheta_c}}{3}.
\]

Particularly, at $\vartheta_c\to 0$ we obtain $F_p = 1+\Re\{r_{bg}\}$ in accordance with Eq.~\eqref{Fp}, at $\vartheta_c\to \pi/2$ we obtain Eq.~\eqref{Gamma:aver:bg:1} [up to a coefficient $4/3$ which follows from Eq.~\eqref{Gamma:aver:1}].

Figure~\ref{fig:St4} demonstrates the results of calculations. The analysis shows that accounting for the in-plane wavevectors of excitons does not strongly affect the results of previous analysis.